\documentclass[aps,pra,epsf,onecolumn,nofootinbib,superscriptaddress,showpacs,preprintnumbers,amsmath,amssymb,floatfix]{revtex4-1}
\usepackage[utf8]{inputenc}
\usepackage[T1]{fontenc}
\usepackage{graphicx}
\usepackage{grffile}
\usepackage{longtable}
\usepackage{wrapfig}
\usepackage{rotating}
\usepackage[normalem]{ulem}
\usepackage{amsmath}
\usepackage{textcomp}
\usepackage{amssymb}
\usepackage{capt-of}
\usepackage{hyperref}
\date{}
\title{Pattern formation in one-dimensional polaron systems// and temporal orthogonality catastrophe}
\hypersetup{
 pdfauthor={Georgios Koutentakis},
 pdftitle={Pattern formation in one-dimensional polaron systems and temporal orthogonality catastrophe},
 pdfkeywords={},
 pdfsubject={},
 pdfcreator={Emacs 27.1 (Org mode 9.3)}, 
 pdflang={English}}
\begin{document}

\title{Pattern formation in one-dimensional polaron systems and temporal orthogonality catastrophe}

\author{G.M. Koutentakis}
\affiliation{Center for Optical Quantum Technologies,
Department of Physics,
University of Hamburg, Luruper Chaussee 149, 22761 Hamburg,
Germany}
\affiliation{The Hamburg Centre for Ultrafast Imaging,
University of Hamburg, Luruper Chaussee 149, 22761 Hamburg,
Germany}
\author{S.I. Mistakidis}
\affiliation{ITAMP,
Center for Astrophysics | Harvard \& Smithsonian,
Cambridge, MA 02138, United States of America}
\affiliation{Center for Optical Quantum Technologies,
Department of Physics,
University of Hamburg, Luruper Chaussee 149, 22761 Hamburg,
Germany}
\author{P. Schmelcher}
\affiliation{Center for Optical Quantum Technologies,
Department of Physics,
University of Hamburg, Luruper Chaussee 149, 22761 Hamburg,
Germany}
\affiliation{The Hamburg Centre for Ultrafast Imaging,
University of Hamburg, Luruper Chaussee 149, 22761 Hamburg,
Germany}

\date{\today}

\begin{abstract}
Recent studies have demonstrated that higher than two-body bath-impurity correlations are not important for quantitatively describing the ground state of the Bose polaron.
Motivated by the above, we employ the so-called Gross Ansatz (GA) approach to unravel the stationary and dynamical properties of the homogeneous one-dimensional Bose-polaron for different impurity momenta and bath-impurity couplings.
We explicate that the character of the equilibrium state crossovers from the quasi-particle Bose polaron regime to the collective-excitation stationary dark-bright soliton for varying impurity momentum and interactions.
Following an interspecies interaction quench the temporal orthogonality catastrophe is identified, provided that bath-impurity interactions are sufficiently stronger than the intraspecies bath ones, thus generalizing the results of the confined case.
This catastrophe originates from the formation of dispersive shock wave structures associated with the zero-range character of the bath-impurity potential.
For initially moving impurities, a momentum transfer process from the impurity to the dispersive shock waves via the exerted drag force is demonstrated, resulting in a final polaronic state with reduced velocity.
Our results clearly demonstrate the crucial role of non-linear excitations for determining the behavior of the one-dimensional Bose polaron.
\end{abstract}

\maketitle

\section{Introduction}
\label{sec:orgddb4635}

Polaronic excitations constitute an ubiquitous class of quasi-particles, incorporating important ramifications in multiple branches of physics \cite{AlexandrovDevreese2010}.
In material science polarons are encountered in several classes of technologically relevant materials, for instance, in He droplets \cite{padmore71_impur_imper_bose_gas,gross62_motion_foreig_bodies_boson_system}, polar \cite{LandauPekar1948,Pekar1947,Pekar1946,feynman55_slow_elect_polar_cryst,froehlich54_elect_lattic_field} or organic \cite{fetherolf20_unific_holst_polar_dynam_disor,fratini09_bandl_motion_mobil_satur_organ_molec_semic,kenkre02_finit_bandw_calcul_charg_carrier} semiconductors and transition metal oxides \cite{verdi17_origin_cross_from_polar_to,moser13_tunab_polar_conduc_anatas}, while their broad relevance stretches even towards biophysics \cite{davydov73_theor_contr_protein_under_their_excit}.
Their formation, properties and interactions are key elements in important phenomena such as the electric conductivity of polymers \cite{mahani17_break_polar_conduc_polym_at,bredas85_polar_bipol_solit_conduc_polym}, the organic magnetoresistance \cite{bobbert07_bipol_mechan_organ_magnet}, the Kondo effect \cite{hewson_1993} and even high-temperature superconductivity \cite{SousChakraborty2018,ChakravertyRanninger1998,Lakhno2019,AlexandrovKabanov1996,Mott1993,lee06_dopin_mott_insul}.
Therefore, it is not surprising that ultracold atoms, being one of the prime platforms for quantum simulation \cite{gross17_quant_simul_with_ultrac_atoms_optic_lattic}, have been employed for studying polaronic structures.
In these systems, two different kinds of polaronic excitations have been experimentally realized to date.
Namely, the Fermi polaron \cite{schirotzek09_obser_fermi_polar_tunab_fermi,kohstall12_metas_coher_repul_polar_stron,scazza17_repul_fermi_polar_reson_mixtur,cetina16_ultraf_many_body_inter_impur,wenz13_from_few_to_many,koschorreck12_attrac_repul_fermi_polar_two_dimen} referring to an impurity interacting with an extensive gas of fermionic atoms, and the Bose polaron \cite{catani12_quant_dynam_impur_one_dimen_bose_gas,spethmann12_dynam_singl_neutr_impur_atoms,skou21_non_equil_quant_dynam_format_bose_polar,yan20_bose_polar_near_quant_critic,hu16_bose_polar_stron_inter_regim,joergensen16_obser_attrac_repul_polar_bose_einst_conden}, where the environment possesses a bosonic character.
Accordingly, these systems have recently been a topic of intense theoretical study in the ultracold community especially regarding their stationary properties \cite{massignan14_polar_dress_molec_itiner_ferrom,schmidt18_univer_many_body_respon_heavy,kalas06_inter_induc_local_impur_trapp,cucchietti06_stron_coupl_polar_dilut_gas,astrakharchik04_motion_heavy_impur_throug_bose_einst_conden,schmidt18_univer_many_body_respon_heavy,GrusdtDemler2015,sun17_visual_efimov_correl_bose_polar,christensen15_quasip_proper_mobil_impur_bose_einst_conden,shi18_impur_induc_multib_reson_bose_gas,levinsen15_impur_bose_einst_conden_efimov_effec,yoshida18_univer_impur_bose_einst_conden,CasteelsVanCauteren2011,TempereCasteels2009,ardila19_analy_bose_polar_acros_reson_inter,ardila16_bose_polar_probl,ardila15_impur_bose_einst_conden,grusdt17_bose_polar_ultrac_atoms_one_dimen,jager20_stron_coupl_bose_polar_one_dimen,will21_polar_inter_bipol_one_dimen,panochko19_mean_field_const_spect_one,jager21_stoch_field_approac_to_quenc,volosniev17_analy_approac_to_bose_polar,guenther21_mobil_impur_bose_einst_conden_orthog_catas,drescher20_theor_reson_inter_impur_bose_einst_conden,takahashi19_bose_polar_spher_trap_poten,schmidt21_self_stabil_bose_polar}.
Lately, it has been argued that the ground state of the Bose polaron can be well-described in terms of a simple Gross-Pitaevskii mean-field type variational approach \cite{jager20_stron_coupl_bose_polar_one_dimen,will21_polar_inter_bipol_one_dimen,panochko19_mean_field_const_spect_one,jager21_stoch_field_approac_to_quenc,volosniev17_analy_approac_to_bose_polar,guenther21_mobil_impur_bose_einst_conden_orthog_catas,drescher20_theor_reson_inter_impur_bose_einst_conden,takahashi19_bose_polar_spher_trap_poten,schmidt21_self_stabil_bose_polar}, herewith referred to as the Gross Ansatz (GA)\footnote{To avoid confusion, since within the common Gross-Pitaevskii equation for an ideal BEC all correlations are neglected, here we will instead adopt the term Gross Ansatz (GA) \cite{gross62_motion_foreig_bodies_boson_system} when referring to the technique employed in Ref. \cite{jager20_stron_coupl_bose_polar_one_dimen,will21_polar_inter_bipol_one_dimen,panochko19_mean_field_const_spect_one,jager21_stoch_field_approac_to_quenc,volosniev17_analy_approac_to_bose_polar,guenther21_mobil_impur_bose_einst_conden_orthog_catas,drescher20_theor_reson_inter_impur_bose_einst_conden,takahashi19_bose_polar_spher_trap_poten,schmidt21_self_stabil_bose_polar}.}. 
The latter neglects all correlations except for the two-body bath-impurity ones. 

The dynamics of the Bose polaron has also been actively explored \cite{mistakidis20_induc_correl_between_impur_one,mistakidis19_effec_approac_to_impur_dynam,mistakidis19_correl_quant_dynam_two_quenc,schecter16_quant_impur,johnson11_impur_trans_throug_stron_inter,cai10_inter_induc_anomal_trans_behav,theel20_entan_assis_tunnel_dynam_impur,mistakidis19_dissip_correl_dynam_movin_impur,lausch18_preth_coolin_dynam_impur_bose_einst_conden,kroenke15_correl_quant_dynam_singl_atom,lausch18_preth_coolin_dynam_impur_bose_einst_conden,theel21_many_body_collis_dynam_impur,bougas21_patter_format_correl_impur_subjec,mukherjee20_pulse_contin_driven_many_body,mistakidis20_pump_probe_spect_bose_polar,mistakidis21_radiof_spect_one_dimen_trapp_bose_polar,katsimiga18_many_body_dissip_flow_confin,mistakidis20_many_body_quant_dynam_induc,mistakidis19_quenc_dynam_orthog_catas_bose_polar}.
Among the many different facets of the polaron dynamics, here we will focus on the phenomenon of temporal orthogonality catastrophe \cite{jager21_stoch_field_approac_to_quenc,mistakidis20_pump_probe_spect_bose_polar,mistakidis21_radiof_spect_one_dimen_trapp_bose_polar,katsimiga18_many_body_dissip_flow_confin,mistakidis20_many_body_quant_dynam_induc,mistakidis19_quenc_dynam_orthog_catas_bose_polar}.
The latter occurs when an impurity is embedded into an adequately strongly repulsive Bose gas and is manifested by the rapid evolution of the system state towards a configuration orthogonal to the initial one, signifying the dynamical decay of the Bose polaron.
In particular, the temporal orthogonality catastrophe has been extensively explored in the case of confined one-dimensional (1D) Bose gases, where an effective potential description, delineated by the bath density and impurity-medium coupling, has been found to be crucial for understanding the dynamical behavior of the system \cite{theel21_many_body_collis_dynam_impur,bougas21_patter_format_correl_impur_subjec,mistakidis21_radiof_spect_one_dimen_trapp_bose_polar,katsimiga18_many_body_dissip_flow_confin,mistakidis20_many_body_quant_dynam_induc,mistakidis19_quenc_dynam_orthog_catas_bose_polar,theel20_entan_assis_tunnel_dynam_impur,boudjemaa20_breat_modes_repul_polar_bose_bose_mixtur,johnson12_breat_oscil_trapp_impur_bose_gas}.
This potential is speculated to be the origin of the temporal orthogonality catastrophe, leading to the question whether a similar mechanism appears in the homogeneous setting where the notion of the effective potential does not exist.
Recent studies indicate that this actually might be the case \cite{jager21_stoch_field_approac_to_quenc,drescher20_theor_reson_inter_impur_bose_einst_conden,guenther21_mobil_impur_bose_einst_conden_orthog_catas}.
One of our central objectives is thus to address this issue and reveal the origin of the temporal orthogonality catastrophe phenomenon for homogeneous Bose gases.

A main culprit for the manifestation of this phenomenon in homogeneous systems refers to the possible emission of non-linear waves by the Bose Einstein Condensate (BEC).
Importantly, over the past decades the Gross-Pitaevskii equation has proven to perfectly describe such non-linear excitations \cite{kevrekidis2007emergent,kevrekidis15_defoc_nonlin_schroed_equat}. 
The relevant ones for the 1D setting refer, among others, to dark-solitons \cite{weller08_exper_obser_oscil_inter_matter,scott98_format_fundam_struc_bose_einst_conden,fursa97_conver_close_coupl_calcul_elect_helium_scatt} and dispersive shock waves \cite{kamchatnov12_gener_disper_shock_waves_by,chang08_format_disper_shock_waves_by,leboeuf01_bose_einst_beams,hakim97_nonlin_schroed_flow_past_obstac_one_dimen}, which have been also realized experimentally \cite{dutton01_obser_quant_shock_waves_creat,theocharis10_multip_atomic_dark_solit_cigar,burger99_dark_solit_bose_einst_conden,mistakidis18_correl_effec_quenc_induc_phase}. 
In addition, numerous recent studies exemplified that these excitations also occur in the presence of interparticle correlations \cite{mistakidis18_correl_effec_quenc_induc_phase,mistakidis19_dissip_correl_dynam_movin_impur,katsimiga17_many_body_quant_dynam_decay,katsimiga17_dark_brigh_solit_dynam_beyon,syrwid17_time_cryst_behav_excit_eigen,delande14_many_body_matter_wave_dark_solit,martin10_quant_therm_effec_dark_solit,mishmash09_quant_many_body_dynam_dark,dziarmaga03_images_dark_solit_deplet_conden}, albeit possessing a more involved behavior than their mean-field counterparts.
In this context, it is crucial to answer whether such non-linear excitations contribute to the dynamics of the Bose polaron, a question which is further mandated by the similarity between the GA equations-of-motion and the Gross-Pitaevskii one.

In this work we employ the GA formulation to examine the equilibrium and dynamical properties of the repulsive Bose polaron and its relation to non-linear pattern formation.
After revisiting the ground state behavior of the Bose polaron \cite{jager20_stron_coupl_bose_polar_one_dimen,will21_polar_inter_bipol_one_dimen,panochko19_mean_field_const_spect_one,volosniev17_analy_approac_to_bose_polar}, we focus on the equilibrium properties of a moving polaron, where we unveil the crossover from the polaronic to a dark-bright soliton regime.
The above indicate a quite intriguing crossover of the impurity state which for weak interspecies repulsions and/or impurity momenta realizes a quasi-particle and in the opposite limit contributes to a collective excitation of the bosonic host.
The comparison of the equilibrium results obtained through GA with the Multi-Layer Multi-Configuration Time-Dependent Hartree method for atomic mixtures (ML-MCTHDX) \cite{cao17_unified_ab_initio_approac_to}, verifies the exceptional accuracy of the former in describing the two-particle interspecies correlations of the system. 
In particular, the GA approach provides in this case almost identical results to the correlated ML-MCTDHX method for the energy, effective mass and bath-impurity correlations of the Bose polaron, while it overestimates the polaronic residue.

We subsequently explore the dynamical response of the system within GA, by employing interspecies interaction quenches from zero to a finite repulsive interaction strength.
Here, the temporal orthogonality catastrophe is exhibited for all initial impurity momenta, as long as the bath-impurity interactions are sufficiently stronger than the intraspecies bath ones, a phenomenon that generalizes the results reported in the confined scenario \cite{mistakidis20_pump_probe_spect_bose_polar,mistakidis21_radiof_spect_one_dimen_trapp_bose_polar,katsimiga18_many_body_dissip_flow_confin,mistakidis20_many_body_quant_dynam_induc,mistakidis19_quenc_dynam_orthog_catas_bose_polar}.
Interestingly, we show for the first time that this mechanism is related to the formation of dispersive shock wave structures associated with the short-range character of the bath-impurity potential.
Note that independently shock wave formation has been demonstrated after the collision of two polaronic clouds immersed in a Fermi medium 
\cite{tajima20_collis_dynam_polar_cloud_immer_fermi_sea}.
In all cases, the post quench state of the system corresponds to a Bose polaron, accompanied by two dispersive shock wave excitations travelling away from the impurity and having a velocity equal to the speed of sound.
For moving impurities, we monitor the drag force being exerted by the bosonic host to the impurity and resulting in a momentum transfer from the impurity to the emitted dispersive shock waves.
This process leads to the final polaronic state possessing a reduced velocity when compared to the initial one and tending to vanish for strong repulsions as a consequence of the amplification of the drag-force.
The above demonstrate the crucial role of non-linear excitations in the dynamics of the Bose polaron.

This work is structured as follows.
Section \ref{sec:hamiltonian} introduces the homogeneous binary mixture setup and the concept of the Lee-Low-Pines transformation. 
In section \ref{sec:gross-ansatz} we present our main theoretical approach in terms of the GA, which we apply to characterize the static and moving Bose polaron at equilibrium.
In order to establish the validity of the GA approach in characterizing the 1D polaron, in Sec. \ref{sec:correlations} we compare our GA results with the fully correlated ML-MCTDHX approach.
The dynamics of the Bose polaron, associated with the emergence of the temporal orthogonality catastrophe phenomenon is outlined in Sec. \ref{sec:toc}. 
In Sec. \ref{sec:conclusions} we summarize our results and provide future perspectives for further study.
Appendix \ref{sec:renorm} elaborates on the bosonic momentum renormalization in the thermodynamic limit of 1D systems, while appendix \ref{sec:potential-appendix} explores the impact of the range of the interspecies interaction potential on the nature of the emitted excitations during the dynamics.
Finally, appendix \ref{sec:mlx} outlines the ingredients of the employed computational approaches.

\section{Polaron Hamiltonian and Lee-Low-Pines Transformation \label{sec:hamiltonian}}
\label{sec:orgb488f23}

We consider a system of \(N_B\) bosons of mass \(m_B\) interacting with a single impurity atom of mass \(m_I\) within an 1D ring of perimeter \(L\).
It is described by the Hamiltonian
\begin{equation}
\label{hamilt}
\begin{split}
    \Hat{H}=&- \frac{\hbar^2}{2 m_{B}} \sum_{k=1}^{N_B} \frac{\partial^2}{\partial x_k^2} - \frac{\hbar^2}{2 m_I} \frac{\partial^2}{\partial x_I^2}
    +g_{BI} \sum_{k=1}^{N_B} \delta(x_k -x_I) + g_{BB} \sum_{k=1}^{N_B} \sum_{k'<k} \delta(x_k -x_{k'}),
\end{split} 
\end{equation}
where \(x_k\), \(k = 1,\dots,N_{B}\), correspond to the coordinates of the bath particles and \(x_I\) refers to the position of the impurity.
In addition, \(g_{BB}\) and \(g_{BI}\) correspond to the intraspecies interactions of the bath atoms and the interspecies coupling among the bath atom and the impurity respectively.
Notice, that ring confinement of ultracold gases is experimentally feasible \cite{bell16_bose_einst_conden_large_time,beattie13_persis_curren_spinor_conden}.
Here we are also interested in the limit \(L \to \infty\), where our results converge to the thermodynamic limit of homogeneous systems and the boundary conditions become irrelevant.
In this context, box potentials emulating the homogeneous thermodynamic limit results can be also realized experimentally \cite{HueckLuick2018,MukherjeeYan2017,CormanChomaz2014,GauntSchmidutz2013}.
In either case, the system is adequately described as 1D when it is subjected to a strong confinement along the transverse spatial directions.
The transverse confinement leads to the modification of the scattering length of the atomic collisions and allows for the control of the involved interaction strengths via confinement and Fano-Feshbach resonances \cite{ChinGrim2010}.

There are several theoretical approaches to tackle the properties of the Hamiltonian of Eq. \eqref{hamilt} for small \(g_{BB}\).
Traditionally they mainly relied on the linearization of the intraspecies interaction term of the bath via the Bogoliubov approach \cite{CasteelsVanCauteren2011,TempereCasteels2009,GrusdtDemler2015,tajima21_polar_probl_ultrac_atoms}.
Here we will take an alternative route based on the spatial homogeneity of the system, which allows us to further simplify the Hamiltonian of Eq. \eqref{hamilt}, by performing the so-called Lee-Low-Pines transformation \cite{gurari53_xxxvi,lee52_motion_slow_elect_polar_cryst,lee53_motion_slow_elect_polar_cryst}.
The latter is a coordinate transformation to the frame-of-reference of the impurity, namely \(r_k=x_k-x_I\) and \(r_I = x_I\).
The transformed Hamiltonian reads
\begin{equation}
    \label{hamilt_LLP}
    \begin{split}
    \hat{H}_\mathrm{LLP} =- \frac{\hbar^2}{2 m_r} \sum_{k=1}^{N_B} \frac{\partial^2}{\partial r_k^2}
    +g_{BI} \sum_{k=1}^{N_B} \delta(r_k) + g_{BB} \sum_{k=1}^{N_B} \sum_{k'<k} \delta(r_k -r_{k'})\\
    - \frac{\hbar^2}{2 m_I} \frac{\partial^2}{\partial r_I^2}
    - \frac{\hbar^2}{m_I} \sum_{k=1}^{N_B} \sum_{k'<k} \frac{\partial }{\partial r_k}\frac{\partial }{\partial r_{k'}}
    + \frac{\hbar^2}{m_I} \sum_{k=1}^{N_B} \frac{\partial }{\partial r_k}\frac{\partial }{\partial r_{I}},
    \end{split}
\end{equation}
where \(m_r = (m_B^{-1} + m_I^{-1})^{-1}\) is the reduced mass of the bath-impurity system.
Interestingly, the momentum operator of the impurity \(\hat{p}_I=-i \hbar \frac{\partial}{\partial r_{I}}\) commutes with the transformed Hamiltonian, Eq. \eqref{hamilt_LLP}, and therefore the momentum of the impurity is conserved in the Lee-Low-Pines transformed frame.
This allows us to simplify the two-species system into an effective single-species one, by replacing \(\hat{p}_{I}=p_{I} \in
\mathbb{R}\).
This reduction comes with the expense of having to deal with an additional momentum-momentum interaction term for the bath atoms, with a coupling inversely proportional to the mass of the impurity \(m_{I}\).

\section{Gross Ansatz treatment of the Lee-Low-Pines Hamiltonian \label{sec:gross-ansatz}}
\label{sec:orga1e923e}

In the case \(m_{I} \to \infty\), \(\hat{H}_{\mathrm{LLP}}\) reduces to the well-studied Lieb-Liniger model \cite{lieb63_exact_analy_inter_bose_gas_I}, with an additional \(\delta\)-shaped potential at the origin \(r=0\).
It is known \cite{lieb63_exact_analy_inter_bose_gas_II} that the excitation spectrum of the Lieb-Liniger model is well described by the Bogoliubov one \cite{PitaevskiiStringari2016} for \(\gamma_{\rm LL}=2 m_B g_{BB}/(\hbar^2 n_0) \ll 1\), where \(n_0\) is the density of the bath atoms, \(n_0 =N_B /L\).
This motivates a mean-field treatment of the Hamiltonian of Eq. \eqref{hamilt_LLP} in the case of small \(g_{BB}\).
In particular, we expand the state of the system in terms of the so-called GA, \(| \Psi (t) \rangle = | \Psi_{GA} (t) \rangle\), with \cite{schurer17_unrav_struc_ultrac_mesos_collin_molec_ions,gross62_motion_foreig_bodies_boson_system},
\begin{equation}
    \label{Gross}
    \Psi_{GA}(r_I,r_1,\dots,r_{N_B};t)=\frac{1}{\sqrt{L}} e^{\frac{i}{\hbar} p_I r_{I}} \prod_{k=1}^{N_B} \psi(r_k;t),
\end{equation}
with \(p_I\) being the momentum of the impurity in the Lee-Low-Pines frame. 
Additionally, \(\psi(r;t)\) is the single particle wavefunction occupied by all the bath atoms.
Note here that, within GA, \(\psi(r;t)\) depends only on \(r_k = x_k - x_I\). 
Another important feature of this wavefunction ansatz, Eq. \eqref{Gross}, is that it neglects all correlations emanating among the bath particles.
As a consequence, it assumes that the bath despite the presence of the impurity, remains in a BEC state.
Nevertheless, the correlations among the impurity and the bath particles are properly taken into account.
This can be verified by considering the two-body density of the bath and the impurity atoms
\begin{equation}
    \label{two_body}
    \rho^{(2)}_{IB}(x_I;x_1;t)=n_0 |\psi(x_1-x_I;t)|^2=n_0 |\psi(r_1;t)|^2 \neq \rho^{(1)}_B(x_1;t) \rho^{(1)}_I(x_I;t) = \frac{n_0}{L}.
\end{equation}
In summary, the GA allows us to obtain the variationally optimal two-body correlations between the impurity and the bath, by neglecting all higher-order correlations \cite{gross62_motion_foreig_bodies_boson_system}.

\subsection{The polaron solution}
\label{sec:orge3fa478}

To find the variationally optimal configuration within the GA approximation we have to minimize the energy functional stemming from \(\hat{H}_{\rm LLP}\), under the constraint of a normalized \(\psi(r)\).
The corresponding functional can be obtained by e.g. following the Dirac-Frenkel variational principle \cite{dirac30_note_exchan_phenom_thomas_atom,Frenkel1934} and introducing the Lagrange coefficient \(\mu(t)\)
\begin{equation}
    \label{functional}
    \begin{split}
        E\left[ \psi(r;t) \right] &= \left\langle \Psi_{GA} (t) \left| \hat{H}_\mathrm{LLP} - i \hbar \frac{\mathrm{d}}{\mathrm{d} t} \right| \Psi_{GA}(t)\right\rangle +\mu(t) N_B \left(1-\int \mathrm{d} r~ |\psi(r;t)|^2 \right)\\
    &=\frac{p_I^2}{2 m_I} + \mu(t) N_B + N_B \int \mathrm{d} r \bigg[-i \hbar \psi^*(r;t) \frac{\partial \psi(r;t)}{\partial t} -\frac{\hbar^2}{2 m_r} \psi^*(r;t) \frac{\partial^2 \psi(r;t)}{\partial r^2}  \\
&+ g_{BI} \delta(r) |\psi(r;t)|^2
        -\mu(t) |\psi(r;t)|^2 + \frac{g_{BB}}{2} (N_B-1) |\psi(r;t)|^4 +\frac{i \hbar p_I}{m_I} \psi^*(r;t) \frac{\partial \psi(r;t)}{\partial r} \\
    &-\frac{\hbar^2 (N_B-1)}{2 m_I} \left( \int \partial r'~\psi^*(r';t) \frac{\partial \psi(r';t)}{\partial r'} \right) \psi^*(r;t) \frac{\partial \psi(r;t)}{\partial r}\bigg].
    \end{split}
\end{equation}
The variation of Eq. \eqref{functional} yields the Gross-Pitaevskii type \cite{PitaevskiiStringari2016} equation
\begin{equation}
    \label{GPE}
    \begin{split}
    i \hbar \frac{\partial}{\partial t} \psi(r;t) = \bigg[-\frac{\hbar^2}{2 m_r} \frac{\partial^2}{\partial r^2} +\frac{i \hbar^2 k_0(t)}{m_I} \frac{\partial}{\partial r} +g_{BI} \delta(r) +g_{BB}(N_B-1) |\psi(r;t)|^2 -\mu \bigg]\psi(r;t),
    \end{split}
\end{equation}
where \(\hbar k_0(t) \equiv  \left(p_I +i\hbar(N_B-1)  \int \mathrm{d}r'~\psi^*(r';t) \frac{\partial \psi(r';t)}{\partial r'} \right)\).
Notice here the non-linear dependence of Eq. \eqref{GPE} on \(\frac{\partial \psi}{\partial r}\), which goes beyond the framework of the standard Gross-Pitaevskii equation and accounts for the coupling of the impurity momentum with the state of the bath.

Herewith, let us focus on stationary solutions, \(\psi(r;t) = \psi(r)\), where Eq. \eqref{functional} reduces to the corresponding energy functional and \(k_0(t) = k_{0}\), \(\mu(t) = \mu\).
We remark that Eq. \eqref{GPE} has already been solved in Ref. \cite{hakim97_nonlin_schroed_flow_past_obstac_one_dimen} for \(N_B,L \to \infty\), while \(n_0 = N_B / L = \text{finite}\) and in the case of a given value of \(k_0\).
Setting \(\psi(r) = \sqrt{n(r)/N_B} e^{i \varphi(r)}\) the ingredients of the underlying solution read
\begin{equation}
    \label{Hakim_solution}
    \begin{split}
    n(r)&=n_0 \left[ \beta^2 + \frac{1}{\gamma^2} \tanh^2 \frac{|r|+r_0}{\sqrt{2}\gamma\xi} \right], \text{  and}\\
    \varphi(r) &=\frac{r}{|r|}\left[ \tan^{-1}\left( \frac{1}{\beta\gamma} \tanh \frac{r_0}{\sqrt{2}\gamma\xi} \right)-
    \tan^{-1}\left( \frac{1}{\beta\gamma} \tanh \frac{|r|+r_0}{\sqrt{2}\gamma\xi}  \right)\right],
    \end{split}
\end{equation}
with \(\beta=v/c\), \(\gamma=(1-\beta^2)^{-\frac{1}{2}}\), the speed of sound defined as \(c=\sqrt{g_{BB} n_0/m_r}\) and the flow velocity \(v=\hbar k_0/m_I\) of the BEC relative to the impurity. 
The healing length is \(\xi=\hbar/\sqrt{2 m_{r} g_{BB} n_0}\) and the Lagrange coefficient, \(\mu =g_{BB} n_0\), which can be identified as the chemical potential \cite{PitaevskiiStringari2016}.
In order to express the solution belonging to Eq. \eqref{Hakim_solution} in terms of the system parameters \(g_{BI}\) and \(p_I\), the values of \(r_0\) and \(v\) have to be determined self-consistently by solving the following two algebraic equations
\begin{equation}
    \label{self_consist}
    \begin{split}
    g_{BI}&=\frac{\hbar c}{\gamma^3}\frac{\tanh \frac{r_0}{\sqrt{2}\gamma\xi}}{\beta^2+\sinh^2 \frac{r_0}{\sqrt{2}\gamma\xi}},\\
    p_I &= \frac{\hbar \beta}{\xi}\left[ -\frac{1}{\sqrt{2}}\frac{m_I}{m_r} + \frac{2 n_0 \xi}{\gamma}\left( 1 - \tanh \frac{r_0}{\sqrt{2}\gamma\xi} \right) \right] -\hbar n_0 \Delta\varphi. 
    \end{split}
\end{equation}
Here \(\Delta\varphi\) is the phase difference of the BEC wavefunction, \(\psi(r)\), at \(r=\pm \infty\), namely
\begin{equation}
    \label{phase_shift}
    \begin{split}
    \Delta\varphi &=\lim_{r \to \infty} \varphi(r) -\lim_{r\to - \infty} \varphi(r)\\
    &=2 \left[ \tan^{-1}\left( \frac{1}{\beta\gamma} \tanh \frac{r_0}{\sqrt{2}\gamma\xi} \right)-
    \tan^{-1}\left( \frac{1}{\beta\gamma} \right)\right].
    \end{split}
\end{equation}
Before proceeding, let us stress that the solution of Eq. \eqref{Hakim_solution} possesses unconventional boundary conditions as the wavefunction changes by a phase factor \(e^{i \Delta \varphi}\) from \(r \to - \infty\) to \(r \to +\infty\).
This is the reason of the existence of the term \(\propto \Delta\varphi\) in Eq. \eqref{self_consist}.
In particular, in the presence of such boundary conditions the bosonic momentum needs to be renormalized by a finite amount \cite{PitaevskiiStringari2016} (see also Appendix \ref{sec:renorm}).
This implies that \(p_I\) is not connected with \(v\) via the relation \(p_I = m_I v\), a fact that will become particularly important later on.
Additionally, a phase difference \(\Delta \varphi \neq 0\) which is realized for \(p_I \neq 0\) implies a global change in the BEC wavefunction \(\psi(r)\).
This feature indicates that the 1D Bose polaron, within GA, possesses the character of a collective excitation of the BEC.

\subsection{The case of a static polaron \label{sec:static-polaron-GA}}
\label{sec:org45fb380}

Recently, the properties of the Gross-Pitaevskii type Eq. \eqref{GPE} have been intensively studied \cite{jager20_stron_coupl_bose_polar_one_dimen,will21_polar_inter_bipol_one_dimen,panochko19_mean_field_const_spect_one,jager21_stoch_field_approac_to_quenc,volosniev17_analy_approac_to_bose_polar,guenther21_mobil_impur_bose_einst_conden_orthog_catas,drescher20_theor_reson_inter_impur_bose_einst_conden,takahashi19_bose_polar_spher_trap_poten} in one and three spatial dimensions.
Below we will briefly review the implications of the 1D solution Eq. \eqref{Hakim_solution} on the properties of the polaron \cite{jager20_stron_coupl_bose_polar_one_dimen,will21_polar_inter_bipol_one_dimen,panochko19_mean_field_const_spect_one,jager21_stoch_field_approac_to_quenc,volosniev17_analy_approac_to_bose_polar}.
For \(p_I = 0\) the self consistency Eq. \eqref{self_consist} can be solved exactly, yielding \(\beta =0\) and 
\begin{equation}
\label{solutionx0}
r_0 = \frac{\xi}{\sqrt{2}}  \sinh^{-1} \left( \frac{2 \hbar c}{g_{BI}} \right).
\end{equation}
Consequently, most of the properties of the Bose polaron depend on the ratio,
\begin{equation}
\label{factor}
\frac{g_{BI}}{2 \hbar c} = \sqrt{\gamma_{\rm LL}} \frac{g_{BI}}{g_{BB}} \left( 1 + \frac{m_B}{m_I} \right)^{-\frac{1}{2}}.
\end{equation}
Recall that the GA description is expected to be valid as long as \(\gamma_{\rm LL} \ll 1\).
Therefore, the behavior of the polaron is mainly tunable via the ratio of the intra and interspecies interaction strengths and the mass imbalance among the impurity and the bath particles.
However, this mass ratio, \(m_B / m_I\), affects Eq. \eqref{factor} only weakly since \(0.18 < \left( 1 + \frac{m_B}{m_I} \right)^{- \frac{1}{2}} < 0.98\) for all currently experimentally realizable ultracold setups\footnote{The lower bound corresponds to a \({}^{6}\)Li impurity immersed in a \({}^{176}\)Yb bath and the upper bound to a \({}^{176}\)Yb impurity immersed in a Bose medium of \({}^7\)Li.},
leading to the conclusion that the most important factor for characterizing the state of the polaron is the interaction strength fraction \(g_{BI} / g_{BB}\).

\begin{figure}[htbp]
\centerline{\includegraphics[width=\textwidth]{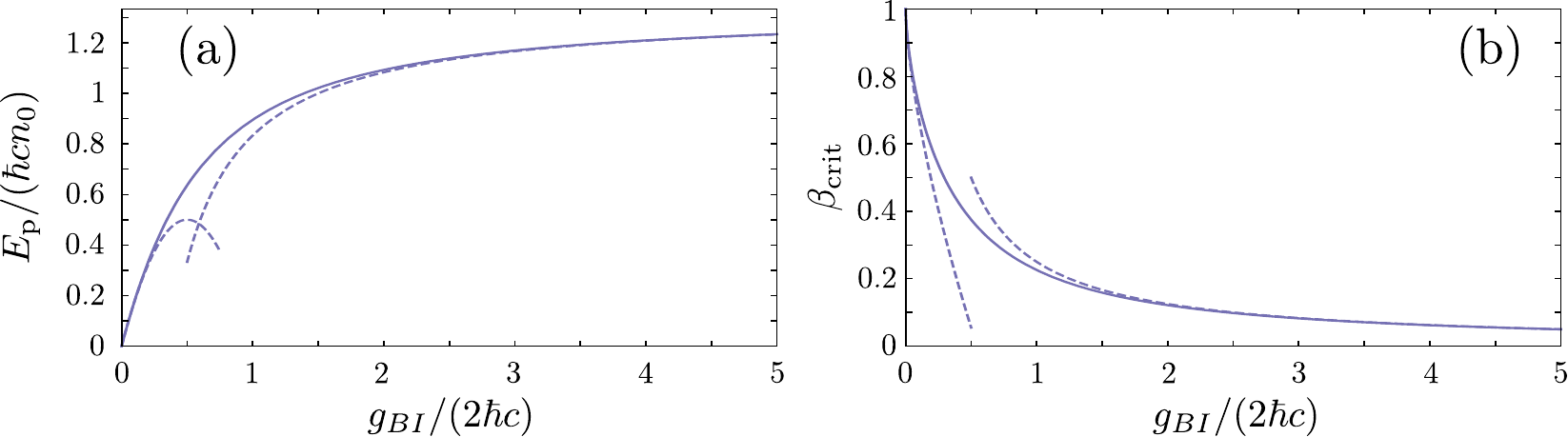}}
\caption{\label{fig:analytics} Analytical predictions for the energy and critical velocity of the Bose polaron within GA. (a) Polaron energy, \(E_{\rm p}\) [Eq. \eqref{polaron_energy}] and (b) critical velocity of the polaronic solution, \(\beta_{\rm crit}\), [Eq. \eqref{critical_velocity}] for varying bath-impurity interaction strength, \(g_{BI}\). In both cases the solid lines indicate the exact results while the dashed lines correspond to weak (leftmost line), \(\mathcal{O}\left( \frac{g_{BI}}{2 \hbar c} \right)^2\), and strong (rightmost line), \(\mathcal{O}\left(\frac{2 \hbar c}{g_{BI}}\right)^2\), asymptotic Taylor expansions.}
\end{figure}

A quantity that will be important for the description of the polaron dynamics is its energy, \(E_{\rm p} = E[ \psi(r) ] -E_{0}\), with \(E_0=g_{BB} n_0^2/2\), being the excess energy of the polaron state with respect to the energy of the system for \(g_{BI} =0\). 
Thus, the energy of the static polaron (see also Eq. \eqref{functional}) reads
\begin{equation}
    \label{polaron_energy}
    E_{\rm p}=\frac{\hbar c n_0}{3}\left\{ 4 - \left[ \sqrt{1+\left( \frac{g_{BI}}{2 \hbar c} \right)^{2}} - \frac{g_{BI}}{2 \hbar c} \right]^3 - 3\left[ \sqrt{1+\left( \frac{g_{BI}}{2 \hbar c} \right)^{2}} - \frac{g_{BI}}{2 \hbar c} \right] \right\}.
\end{equation}
A simple Taylor expansion in terms of \(\frac{g_{BI}}{2 \hbar c}\), demonstrates that the energy of the polaron within GA grows linearly for small \(g_{BI}\), as is also expected for the non-interacting BEC background, \(\psi(r) = \sqrt{n_0}\).
Significant deviations only appear when \(g_{BI} \approx 2 \hbar c\), where the energy of the polaron becomes smaller than the one of the corresponding non-interacting profile, \(E_{{\rm NI}} = g_{BI} n_0\), since the BEC density in the vicinity of the impurity is suppressed.
For strong repulsions, namely \(g_{BI} / (2 \hbar c)  \gg 1\), the energy of the polaron saturates to the value \(4 \hbar c n_0 /3\), a tendency which has been shown to qualitatively agree with corresponding Quantum Monte Carlo predictions in Ref. \cite{grusdt17_bose_polar_ultrac_atoms_one_dimen}.

Figure \ref{fig:analytics} (a), demonstrates the behavior of \(E_{\rm p}\) over the characteristic energy scale \(\hbar c n_0\) for increasing \(g_{BI} / (2 \hbar c)\).
By comparing the behavior of this quantity with the first order asymptotics of Eq. \eqref{polaron_energy}, we can observe the emergence of the three distinct interaction regimes.
Namely, for small, \(g_{BI} / (2 \hbar c) < 0.25\), and large values, \(g_{BI} / (2 \hbar c) > 2\), the  impurity-medium energy, \(E_{\rm p}\), matches the results of the corresponding asymptotic expansions.
In contrast, within the intermediate interaction regime \(0.25 < g_{BI} / (2 \hbar c) < 2\)  deviations between the exact values of \(E_{\rm p}\), Eq. \eqref{polaron_energy}, and the approximate Taylor expansions occur.
Finally, let us note that the typical energy scale of the system \(\hbar c n_0\) is related to the corresponding interaction-independent one \(\hbar^2 n_0^{2} /m_B\) via
\begin{equation}
\label{soliton-energy-scale}
\hbar c n_0 = \sqrt{\frac{\gamma_{\rm LL}}{2} \left( 1 + \frac{m_B}{m_I}\right) } \frac{\hbar^2 n_0^2}{m_B},
\end{equation}
which is a function of the Lieb-Liniger parameter, \(\gamma_{\rm LL}\), and the mass ratio, \(m_B / m_I\).
The above indicate that the energy scale of the Bose polaron is small compared to the non-interacting energy scale, \(\hbar^2 n_0^2 / m_B\), at least when we focus on the case of a BEC host in which \(\gamma_{\rm LL} \ll 1\).

\subsection{Moving polaron and the soliton solution \label{sec:moving-polaronGA}}
\label{sec:org847f6c6}

Having briefly commented on the analytic polaron solution for \(p_I = 0\), let us elaborate on the case of a moving polaron with \(p_I \neq 0\), where no analytic solution exists and it has been far less discussed in the literature.
In that case, the parameters of the polaron need to be found numerically by solving the self-consistency Eq. \eqref{self_consist}, for \(r_0\) and \(\beta\) \cite{panochko19_mean_field_const_spect_one}.
It can be easily proven that solutions of Eq. \eqref{self_consist} for \(r_0\) exist only in the case that the velocity of the polaron \(\beta\), does not exceed the critical one \(\beta_{\rm crit}\) \cite{hakim97_nonlin_schroed_flow_past_obstac_one_dimen}.
The value of the critical velocity, \(v_{\rm crit} = \beta_{\rm crit} c\), can be obtained by finding the maximum with respect to \(r_0\) of the right-hand side of the first self consistency equation, Eq. \eqref{self_consist}, see also Ref. \cite{hakim97_nonlin_schroed_flow_past_obstac_one_dimen}.
This process yields the following algebraic equation 
\begin{equation} \label{critical_velocity}
\frac{g_{BI}}{2\hbar c} =  \sqrt{2} \left( 1 - \beta_{\rm crit}^2 \right) \frac{\sqrt{\sqrt{1 + 8 \beta_{\rm crit}^2} - \left(1 + 2 \beta_{\rm crit}^2 \right)}}{4 \beta_{\rm crit}^2 -1 + \sqrt{1 + 8 \beta_{\rm crit}^2}} .
\end{equation}
Notice here, that the value of the critical velocity depends only on the ratio \(g_{BI} / ( 2 \hbar c)\).
The behavior of \(\beta_{\rm crit}\) along with its strong and weak asymptotics is provided in Fig. \ref{fig:analytics} (b).
It can be seen that for \(g_{BI} = 0\) the critical velocity is equal to the speed of sound, \(\beta_{\rm crit} = 1\), and for increasing \(g_{BI}\) it is suppressed.
For large \(g_{BI}\), \(\beta_{\rm crit}\) behaves as \(\beta_{\rm crit} \propto 1/g_{BI}\), in agreement with the predictions of Ref. \cite{hakim97_nonlin_schroed_flow_past_obstac_one_dimen}.

\begin{figure}[h] 
\centering \includegraphics[width=1.0\textwidth]{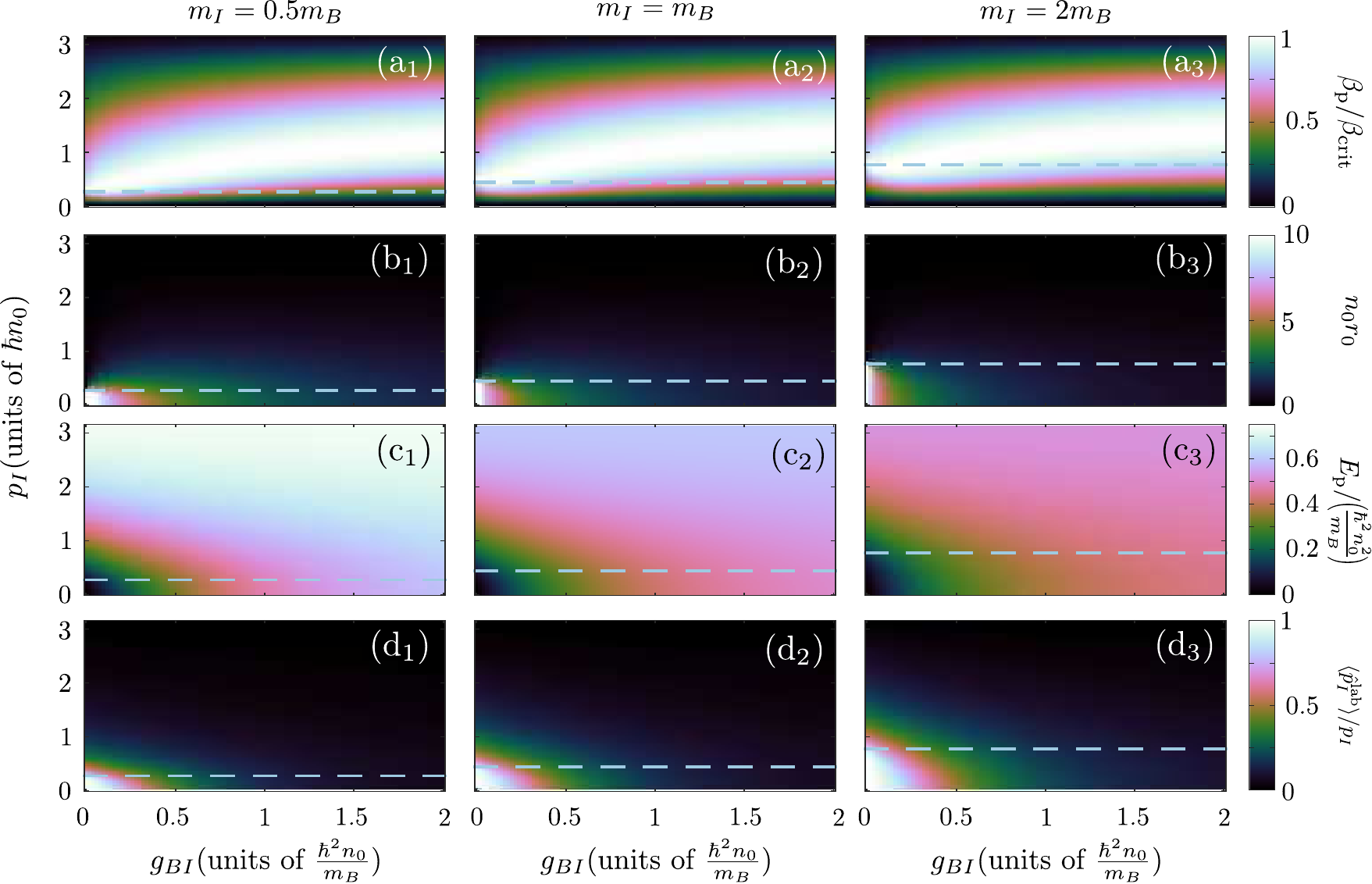} 
\caption{\label{fig:ground_state} Characteristic properties of the moving Bose polaron within GA. (a\({}_1\))-(a\({}_3\)) Velocity of the polaron over its critical one, \(\beta_{\rm p}/\beta_{\rm crit}\), (b\({}_1\))-(b\({}_3\)) offset parameter, \(r_0\), of the polaron solution, (c\({}_1\))-(c\({}_3\)) polaron energy, \(E_{\rm p}\) and (d\({}_1\))-(d\({}_3\)) expectation value ratio of the impurity momentum between the laboratory and the impurity frames, \(\langle \hat{p}^{\rm lab}_I \rangle/p_I\), for different values of \(g_{BI}\) and \(p_I\). The distinct columns correspond to different impurity masses,  \(m_I = m_B /2\) (left panels), \(m_I = m_B\) (middle panels) and \(m_I = 2 m_B\) (right panels). In all cases, the data correspond to thermodynamic limit calculations, \(N,L \to \infty\), with \(g_{BB} = 0.1 \hbar^2 n_0 /m_B\) and dashed lines represent \(p_I = m_I c\).}
\end{figure} 

At a first glance one would expect that the value of \(\beta_{\rm crit}\) can be employed for deriving an upper bound for the maximally allowed \(p_I\), however here we will argue that this is not the case.
In particular, for \(p_I = \pm \hbar n_0 \pi\) the Gross-Pitaevskii type equation \eqref{GPE} can be solved analytically yielding the black soliton solution
\begin{equation} \label{black_soliton}
\psi_{p_I = \pm \hbar n_0 \pi}(r) = \mp \sqrt{n_0} \tanh \frac{r}{\sqrt{2} \xi}.
\end{equation}
Since the impurity lies at \(x_I =0\) being the notch of the black-soliton, this solution actually corresponds to a dark-bright soliton for the composite system.
It might seem contradictory that in the case of relatively large momenta, \(|p_I| = \pi \hbar n_{0}\), a stationary BEC flow is encountered.
However, this counter-intuitive result can be attributed to the fact that \(p_I\) does not refer to the momentum of the impurity in the laboratory frame.
In particular by inverting the Lee-Low-Pines transformation we obtain
\begin{equation} \label{momentum-impurity-lab}
\hat{p}^{\rm lab}_I = p_I - \underbrace{\sum_{i=1}^{N_B} \left( - i \hbar \frac{\partial}{\partial r_i} \right)}_{\equiv \hat{p}_B},
\end{equation}
with \(\hat{p}_B\) being the momentum of the bath in the impurity frame, which is invariant under the frame transformation, since \(\partial/\partial r_i = \partial/\partial x_i\).
Therefore, the conservation law of \(p_I\) in the impurity frame implies that the total momentum, \(p_I = \langle \hat{p}_I^{\rm lab} \rangle + \langle \hat{p}_B \rangle\), is conserved in the laboratory frame.
Notice that for the black soliton solution, \(\langle \hat{p}_B \rangle = \pm \hbar n_0 \pi = p_I\), which agrees with the fact that the solution is static, \(\langle \hat{p}_I^{\rm lab} \rangle =0\).

In order to analyze the crossover from the static polaronic to the black soliton solution, we focus on the solutions of Eq. \eqref{self_consist} for \(0 \le p_I \le \pi\).
The solutions for varying \(p_I\) and \(g_{BI}\) while keeping \(g_{BB} = 0.1 \hbar^2 n_0^2 / m_B\) fixed are presented for  \(m_I = 0.5 m_B\) (Fig. \ref{fig:ground_state} (a\(_{1}\)) - (d\(_{1}\))), \(m_I = m_B\) (Fig. \ref{fig:ground_state} (a\(_{2}\)) - (d\(_{2}\))) and \(m_I = 2 m_B\) (Fig. \ref{fig:ground_state} (a\(_{3}\)) - (d\(_{3}\))).
Independently of the impurity mass the velocity of the polaron satisfies \(\beta_{\rm p} \leq \beta_{\rm crit}\) (Fig. \ref{fig:ground_state} (a\(_i\)), \(i=1,2,3\)), hinting towards the conclusion that the state described by Eq. \eqref{Hakim_solution} is stable\footnote{Importantly, by explicitly evaluating the Hessian matrix for the numerical solutions presented in Fig. \ref{fig:ground_state} we can prove that \(\left( H_{E_{\rm p}} \right)_{ij} = \frac{\partial^2 E_{\rm p}}{\partial a_i \partial a_j}\) is positive definite , where \(i,j = 1,2\), with \(a_1= \beta\), \(a_2=r_0\). This supports the stability of the solution within the subspace spanned by Eq. \eqref{Hakim_solution}.} for every \(p_I\) and \(g_{BI}\).
Moreover, the polaron velocity \(\beta\) exhibits a non-monotonic behavior since for small momenta \(p_I < m_I c\), \(\beta\) is increasing with \(p_I\) until it reaches a maximum at a \(g_{BI}\)-dependent momentum value \(p_{I,0} \geq m_I c\).
Beyond that point \(\beta\) decreases with increasing \(p_I\) until it reaches the value of \(\beta =0\) for \(p_I = \pi\hbar n_0\).
In addition, it can be seen that the solution for \(r_0\) (Fig. \ref{fig:ground_state} (b\(_i\)), \(i=1,2,3\)) is appreciably larger than \(0\) only for \(p_I < m_I c\) (see dashed line) and for \(g_{BI} < 0.5\).

The energy of the moving polaron is presented in Fig. \ref{fig:ground_state} (c\(_i\)), \(i=1,2,3\).
Here there are two notable effects.
For a fixed \(g_{BI}\), \(E_{\rm p}\) depends more weakly on \(p_I\) as the value of \(g_{BI}\) increases.
This is a manifestation of the increase of the effective mass, \(m^{*} = \left( \frac{\partial^2 E_p}{\partial p_I^{2}} \right)^{-1}\), of the polaron with \(g_{BI}\) reported in Ref. \cite{panochko19_mean_field_const_spect_one}.
Moreover, for momenta \(p_I \to \pi\) or large interactions the energy of the polaron tends to saturate to the corresponding energy of the dark-bright soliton solution, Eq. \eqref{black_soliton}, \(E_{\rm b} = \lim_{g_{BI} \to \infty} E_{\rm p} = \frac{4}{3} \hbar n_0 c\).
It is evident from Fig. \ref{fig:ground_state} (c\(_i\)), that this asymptotic value of energy decreases with increasing \(m_I\) a fact that can be understood by inspecting Eq. \eqref{soliton-energy-scale}.

The above indicate two distinct regimes for the behavior of the system, the polaron regime encountered for low momenta \(p_I \sim 0\) and interactions, \(g_{BI} \sim 0\), and the dark-bright soliton regime for high momenta, \(p_I \sim \pm \hbar n_0 \pi\) and/or strong interactions, \(g_{BI} \to \infty\).
To characterize the crossover of these two regimes we employ the quantity \(\langle \hat{p}_I^{\rm lab} \rangle/p_I \in [0,1]\), see Fig. \ref{fig:ground_state} (d\(_i\)) with \(i=1,2,3\).
This quantity compares the momentum contribution of the motion of the impurity, \(\langle \hat{p}_I^{\rm lab} \rangle\), to the induced BEC flow, \(\langle \hat{p}_B \rangle = p_I - \langle \hat{p}_I^{\rm lab} \rangle\). 
Therefore, values proximal to \(1\) indicate that the impurity motion is the dominant contribution and as a consequence the system is in the polaron regime.
On the other hand, values close to \(0\) signify that the dominant contribution is the BEC flow and accordingly the system behaves as a dark-bright soliton.
As Fig. \ref{fig:ground_state} (d\(_i\)) testifies, the polaron regime occurs only for \(p_I < m_I c\) and \(g_{BI} < 0.5\), where also \(r_0 \gg 0\), see also Fig. \ref{fig:ground_state} (a\(_{i}\)). 
Otherwise the state of the system lies within the dark-bright soliton regime.

As already mentioned previously the mass of the impurity does not significantly alter the behavior of the system.
It only affects the system quantitatively by shifting the threshold \(m_I c^2\), where the velocity of the impurity becomes supersonic in the case of \(g_{BI} = 0\).
Indeed, this threshold is related with the crossover between the polaronic and dark-bright soliton regimes causing a shift along \(p_I\) for the structures manifested among the different observables.

Concluding, we are in position to infer the dual character of the 1D polaron as captured by the GA approximation.
For varying interaction strengths and momenta the character of an impurity changes.
In the case of small \(g_{BI}\) and \(p_I\) the impurity contributes to a well-defined quasiparticle associated with the local excitation of its BEC environment due to its presence.
In the opposite scenario of large \(g_{BI}\) or \(p_I \to \pm \pi \hbar n_0\) the impurity is embedded within a collective excitation akin to a static dark-bright soliton.
The exploration of this crossover should provide an interesting perspective for future experiments.

\section{Impact of Correlations and validity of the GA approximation \label{sec:correlations}}
\label{sec:org78a0da0}

Let us now investigate the impact of correlations on the above mentioned properties of the polaron.
For this purpose we employ the Multi-Configuration Time-Dependent Hartree Method for Bosons (MCTDHB) \cite{alon08_multic_time_depen_hartr_method_boson,alon07_unified_view_multic_time_propag}, being a reduction of the ML-MCTDHX \cite{cao17_unified_ab_initio_approac_to}, that is able to capture all the relevant correlations emanating in the system.
Since currently the MCTDHB method can only simulate systems with a definite number of particles, here we will focus (in both the ML-MCTDHB and GA case) on a system with \(N_B=100\) confined within a ring of length \(L = 100 n_0^{-1}\).

Notice that within MCTDHB we can work with the Lee-Low-Pines transformed Hamiltonian of Eq. \eqref{hamilt_LLP} by exploiting the \(p_I\) symmetry of the Hamiltonian of Eq. \eqref{hamilt}.
The many-body wavefunction of the system in this case can be reduced, without any loss of generality, to
\begin{equation}
\label{wavefunction_MCTDH_LLP}
\Psi(x_I,x_1,x_2,\dots,x_{N_B}) = \frac{1}{\sqrt{L}} e^{\frac{i}{\hbar} p_I x_I} \Psi_B\bigg(\underbrace{x_1-x_I}_{r_1},\underbrace{x_2-x_I}_{r_2},\dots,\underbrace{x_{N_B}-x_I}_{r_{N_B}} \bigg).
\end{equation}
Then MCTDHB can be employed in order to variationally optimize the \(\Psi_B (r_1,\dots,r_{N_B})\) part of the many-body wavefunction in the absence of any approximation.
This allows us to probe the effect of correlations between the atoms of the bath that the GA of Eq. \eqref{Gross}, neglects.
For more details regarding our MCTDHB calculations see Appendix \ref{sec:mlx}.

\begin{figure}[h]
    \centering
    \includegraphics[width=1.0\textwidth]{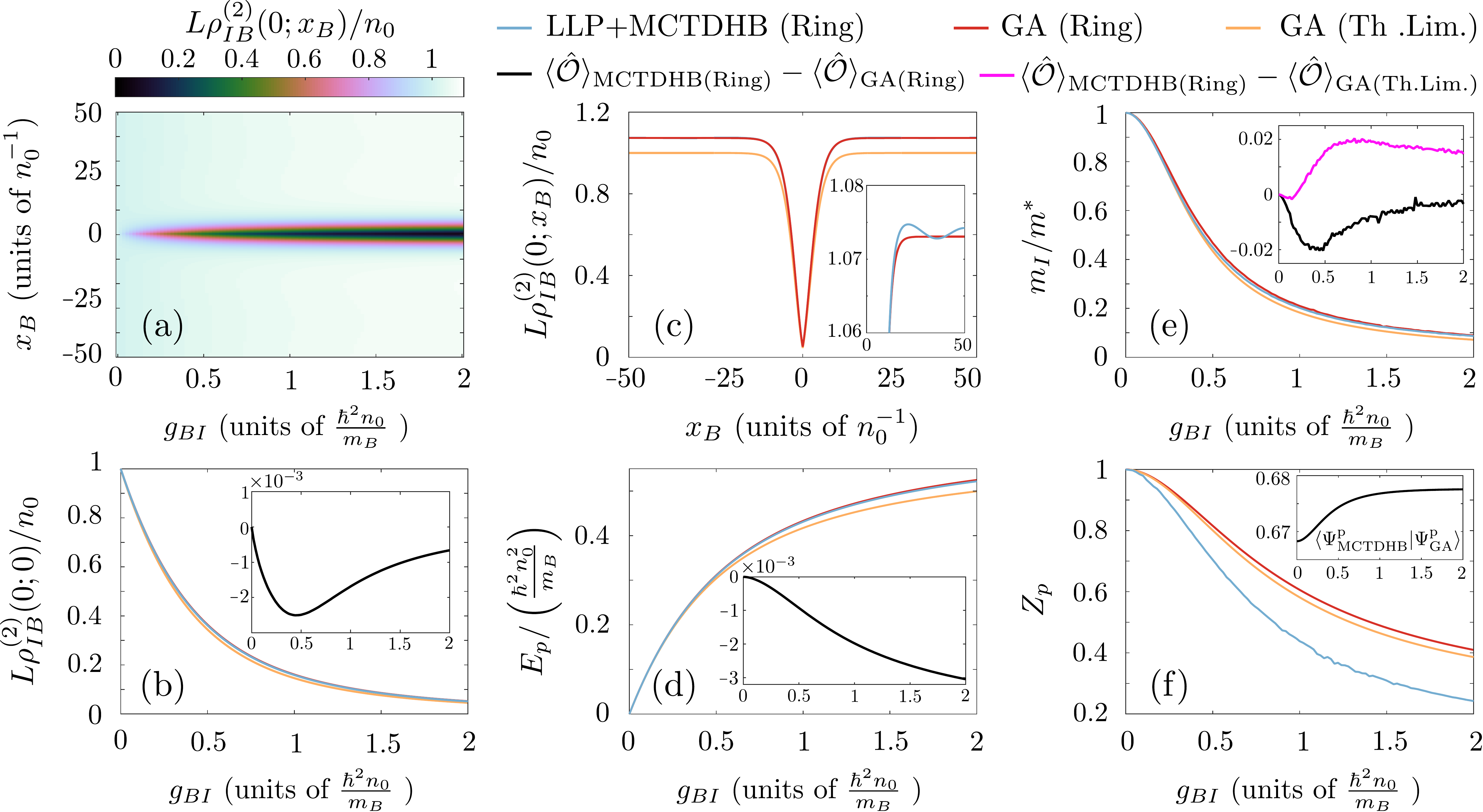}
\caption{\label{fig:comparisonMFMB} Comparison of the GA polaron with the correlated MCTDHB results. (a) Bath-impurity correlations, \(\rho^{(2)}_{IB}(0;x_B)\), for varying interspecies interaction strength, \(g_{BI}\), within MCTHDB. (b) Bath-impurity correlations at coincidence, \(\rho^{(2)}_{IB}(0;0)\), for different \(g_{BI}\) and for all employed approaches (see legend). (c) Comparison of the correlation profile \(\rho^{(2)}_{IB}(0;x_B)\) within the MCTDHB and the GA for \(g_{BI} = 2 \hbar^2 n_0 /m_B\). The inset of (c) provides a magnification of \(\rho_{IB}^{(2)}(0;x_{B})\), showing the behavior of the system away from the impurity. Comparison of (d) the polaron energy, \(E_{\rm p}\), (e) the inverse effective mass, \(m_I/m^{*}\) and (f) the polaron residue among the different approaches and for varying \(g_{BI}\). To elucidate the comparison between GA and MCTDHB, the insets of (b), (d) and (e) provide the difference of the corresponding observables between the distinct approaches (see legend). The inset of (f) indicates the many-body overlap between the MCTDHB and the GA many-body states for varying \(g_{BI}\). In all cases, \(m_I =m_B\), \(p_I =0\) and \(g_{BB} = 0.1 \hbar^2 n_0 /m_B\). The relevant ring confined setups are characterized by \(N_B =100\) and \(L = 100 n_0^{-1}\).}
\end{figure}

The equilibrium state properties of the polaron for weak intraspecies bath repulsions, \(g_{BB}=0.1 \frac{\hbar^2 n_0}{m_B}\), in the equal mass case, \(m_I=m_B\) are compared in Fig. \ref{fig:comparisonMFMB} within the results of the MCTDHB and the GA approaches.
In order to contrast the state of a system confined in a ring of \(L = 100 n_0^{-1}\) to the thermodynamic limit \(N_B, L \to \infty\), we also provide the corresponding results of the GA when extrapolated to the thermodynamic limit, see Ref. \cite{jager20_stron_coupl_bose_polar_one_dimen,will21_polar_inter_bipol_one_dimen,panochko19_mean_field_const_spect_one,jager21_stoch_field_approac_to_quenc,volosniev17_analy_approac_to_bose_polar}. 

The behaviour of the bath-impurity two-body correlations, \(g^{(2)}_{IB}(0;x_B) = \frac{L}{n_0} \rho^{(2)}_{IB}(0;x_B)\), Eq. \eqref{two_body}, for different interspecies repulsions \(g_{BI}\) is demonstrated within the MCTDHB approach in Fig. \ref{fig:comparisonMFMB} (a).
For increasing \(g_{BI}\) we observe a gradual depletion of \(\rho_{IB}^{(2)}(0;x_B)\) in the vicinity of the impurity, \(x_B \approx x_I = 0\), stemming from the repulsive bath-impurity coupling \cite{mistakidis20_many_body_quant_dynam_induc,mistakidis19_dissip_correl_dynam_movin_impur}.
These anti-correlations are accompanied by bunching of bath-impurity correlations, \(\rho_{IB}^{(2)}(0;x_B)>1\), for \(x_B > 10\) (hardly visible in Fig. \ref{fig:comparisonMFMB} (a)) which originates from the conservation of the total particle number of bath atoms on the ring.
To contrast our findings with the approximate GA method, in Fig. \ref{fig:comparisonMFMB} (b) we compare the effect of anticorrelations between the bath and the impurity atoms captured by \(\rho^{(2)}_{IB}(0;0)\) among the different approaches.
Note here, that this quantity is closely related to the Tan contact \cite{volosniev17_analy_approac_to_bose_polar,santana19_scalin_proper_tans_contac,olshanii03_short_distan_correl_proper_lieb,werner12_gener_relat_quant_gases_two_three_dimen,pricoupenko11_isotr_contac_forces_arbit_repres,tan08_gener_virial_theor_press_relat,tan08_large_momen_part_stron_correl_fermi_gas,tan08_energ_stron_correl_fermi_gas}.
We find that the GA is able to reproduce the behavior observed within the MCTDHB approach.
The fully correlated approach predicts only a slightly more pronounced anti-bunching as shown in the inset of Fig. \ref{fig:comparisonMFMB} (b).
Notice also that the results of the ring confined setups agree very well with the thermodynamic limit ones, indicating the insignificance of finite-size effects for \(\rho^{(2)}_{IB}(0;0)\).
To appreciate better the effect of the correlations and the confinement of the particles in a ring, Fig. \ref{fig:comparisonMFMB} (c) compares the bath-impurity correlations, \(\rho^{(2)}_{IB}(0,x_B)\), for strong repulsions, \(g_{BI}=2\), between the two approaches.
As it can be easily deduced the GA results closely follow the MCTDHB ones.
The only deviations occur away from the position of the impurity \(x_B > 10\) (see also the inset of Fig. \ref{fig:comparisonMFMB} (c)), where in the fully correlated case spatial oscillations of the \(\rho^{(2)}_{IB}(0;x_B)\) profile are observed.
These deviations can be explained by the fact that in the bath a correlation hole appears for two atoms being in close proximity (not shown here for brevity, see also Ref. \cite{BrouzosPhD} and references therein).
Notice also that the results referring to the ring geometry yield \(\rho^{(2)}_{IB}(0,\pm L/2) > n_0\). 
This is a consequence of the particle conservation, occurring in order to accustom for the lower density in the vicinity of the impurity.
In contrast, this behavior is absent in the thermodynamic limit, where \(\rho^{(2)}_{IB}(0,|r|> \xi) \approx n_0\).

Figure \ref{fig:comparisonMFMB} (d) reveals that the inclusion of bath-bath correlations does not significantly affect the energy of the polaron \(E_{\rm p}=E(g_{BI}) - E(g_{BI}=0)\).
This leads to the conclusion that the GA provides an excellent prediction for the polaronic energy, in agreement with Ref. \cite{jager20_stron_coupl_bose_polar_one_dimen}.
Of course, the presence of higher-order correlations within the MCTDHB approach results in a slight reduction of the polaronic energy as illustrated in the inset of Fig. \ref{fig:comparisonMFMB} (d).
In contrast, the effect of the ring confinement provides a more important kinetic energy penalty\footnote{The bath density is expelled from the vicinity of the impurity and accumulates in the spatial region away from it \(x_B > 10\), see also Fig. \ref{fig:comparisonMFMB} (c). This density increase leads to a greater kinetic energy scaling quadratically with the bath density. However, since the number of expelled atoms is roughly constant as the perimeter of the ring \(L\) increases this correction becomes negligible for \(L \to \infty\).}, which can be identified by comparing the confined results to the thermodynamic limit case.

Regarding the effective mass, \(m^{*}\), depicted in Fig. \ref{fig:comparisonMFMB} (e), also a remarkable agreement among both methods and system sizes is observed, see in particular the inset of Fig. \ref{fig:comparisonMFMB} (e).
Indeed, the effective mass of the polaron is related to the local correlations of the dressing cloud in the vicinity of the impurity \cite{massignan14_polar_dress_molec_itiner_ferrom}, where finite-size effects are insignificant, and are well described by the GA.
The last quantity of interest for the Bose-polaron is its quasi-particle residue\footnote{Note that the residue of polaronic quasiparticles can be monitored experimentally via radiofrequency spectroscopy \cite{mistakidis21_radiof_spect_one_dimen_trapp_bose_polar,kohstall12_metas_coher_repul_polar_stron}.}, \(Z_{\rm p} = \sqrt{| \langle \Psi_0 | \Psi_{\rm p} \rangle |^2}\), with \(| \Psi_0 \rangle\), \(| \Psi_{\rm p} \rangle\) being the non-interacting and the polaronic states respectively \cite{mistakidis19_quenc_dynam_orthog_catas_bose_polar,mistakidis20_many_body_quant_dynam_induc,mistakidis21_radiof_spect_one_dimen_trapp_bose_polar}.
In Fig. \ref{fig:comparisonMFMB} (f) it can be seen that the results including beyond two-body bath-impurity correlations differ significantly from the GA ones.
This is because the residue, \(Z_{\rm p}\), is related with the many-body wavefunction overlap of the polaronic state to the non-interacting one and therefore correlations of all orders significantly affect this quantity.
In particular, it can be verified that, while finite-size corrections seem to not be significant in the case of the GA, the presence of higher-order correlations suppresses appreciably the polaronic residue.
Importantly, despite the remarkable agreement on the level of two-body correlations the overlap between the MCTDHB and the GA wavefunction for the system confined in the ring ranges from \(67\) to \(68 \%\), see the inset of Fig. \ref{fig:comparisonMFMB} (f).
This reduction of the many-body wavefunction overlap can be explained by the fact that MCTDHB in contrast to the GA allows for the depletion of the bath wavefunction \(\Psi_B (r_1,\dots,r_{N_B})\) due to the presence of quantum fluctuations.
Notice that any depleted many-body state, where even a single bath atom is in a state orthogonal to the BEC wavefunction, \(\psi(r)\), has exactly zero overlap with the fully condensed many-body wavefunction described by the GA.
Nevertheless, in this case, despite the zero overlap of these two many-body states, the corresponding low-order correlation functions would be almost identical for \(N_B \gg 1\).

\section{Dynamical response of the system: the temporal orthogonality catastrophe \label{sec:toc}}
\label{sec:orgf982ae4}

Having identified the main properties of the equilibrium state of the Bose polaron in free 1D space now we proceed by considering its dynamical response.
In particular, in the same manner as in Ref. \cite{mistakidis20_pump_probe_spect_bose_polar,mistakidis20_many_body_quant_dynam_induc,mistakidis19_correl_quant_dynam_two_quenc,mistakidis19_quenc_dynam_orthog_catas_bose_polar,mistakidis19_repul_fermi_polar_their_induc,jager21_stoch_field_approac_to_quenc}, we examine the polaron generation after an abrupt quench of the interaction strength, \(g_{BI}\), from \(g_{BI} = 0\) to some final positive value \(g_{BI}^f >0\).
Within the GA approximation, Eq. \eqref{Gross}, and for \(g_{BI} =0\) the lowest in energy wavefunction of the composite system with a given value of impurity momentum, \(p_I\), reads
\begin{equation}
\label{initial_state}
\Psi_{0}(x_I,x_1,\dots,x_{N_B};p_I) = L^{-\frac{N_B+1}{2}} \exp \left( - \frac{i}{\hbar} p_I x_I \right).
\end{equation}
We are especially interested in observing the overlap of the time-evolved interacting state, \(| \Psi(t) \rangle\) to the initial non-interacting one \(| \Psi(0) \rangle = | \Psi_0 \rangle\).
As already discussed in Ref. \cite{knap12_time_depen_impur_ultrac_fermion,cetina16_ultraf_many_body_inter_impur,mistakidis19_quenc_dynam_orthog_catas_bose_polar} this observable can be directly probed in spectroscopic experiments and allows to address the polaronic properties.
Another, important concept in homogeneous systems is the influence of the impurity momentum on the subsequent dynamics of the quenched, bath-impurity system, which we also consider below.

Regarding the numerical details of our simulations, we have considered \(N_B = 1600\) particles, with \(m_B = m_I\) confined in a ring with perimeter \(L = 1600 n_0^{-1}\).
The increase of the ring perimeter is essential for approaching the thermodynamic limit and avoiding effects stemming from the imposed boundary conditions.
In the following we have exclusively employed the GA approach, since for such large particle numbers ensuring in general the convergence of fully correlated approaches is computationally beyond reach.
Finally, note that all of the numerical values provided below are provided in units of \(\hbar = m_B = n_0 = 1\).

\subsection{Dynamics of a subsonic impurity}
\label{sec:orga9a396c}

\subsubsection{Dynamics of two-body correlations}
\label{sec:org10f453f}

\begin{figure}[h]
    \centering
    \includegraphics[width=1.0\textwidth]{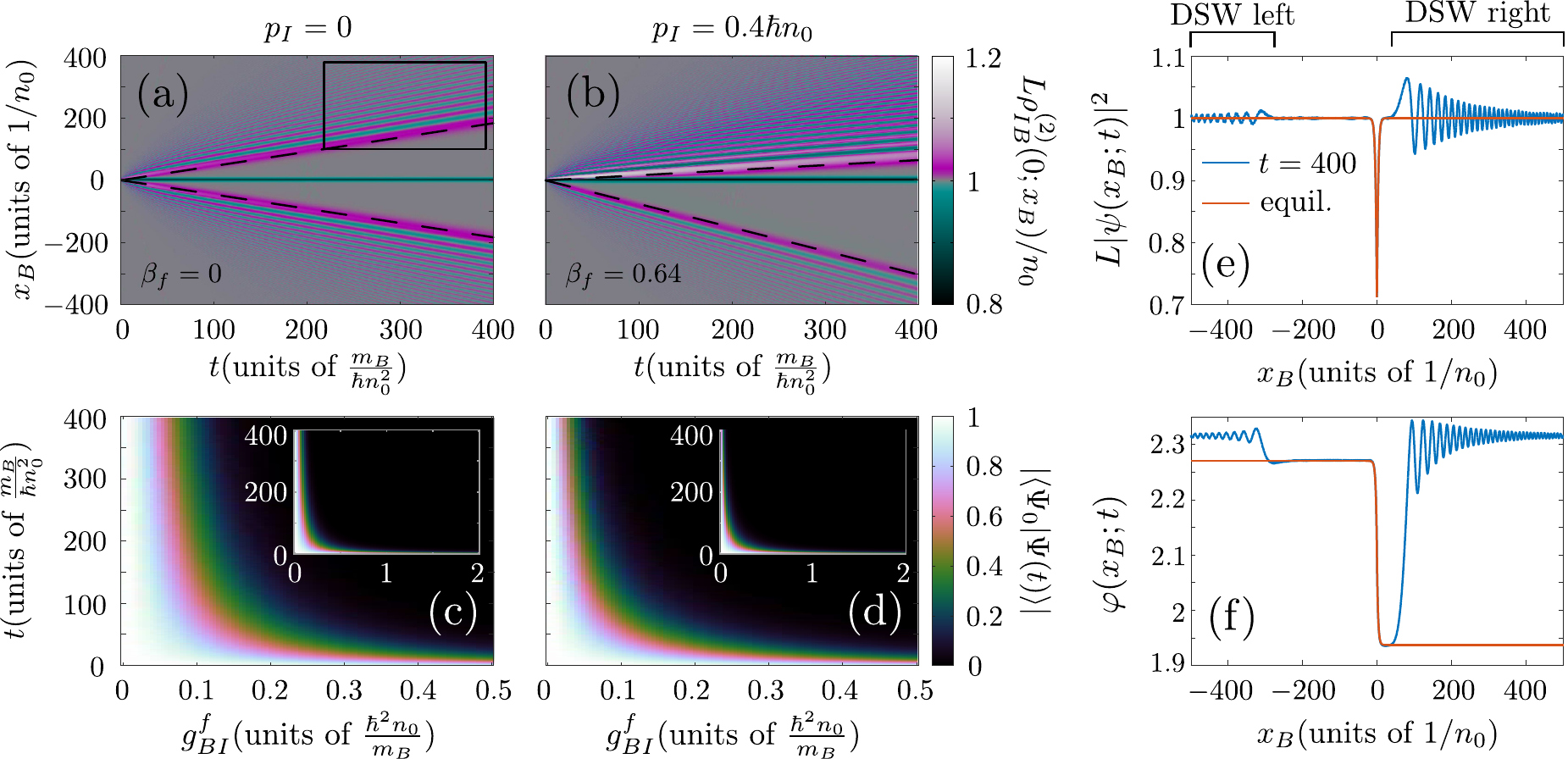}
\caption{\label{fig:dynamicsFid} Quench dynamics of an impurity in a homogeneous Bose gas. (a), (b) Spatiotemporal evolution of the two-body interspecies correlations, \(\rho^{(2)}_{IB}(0;x_B)\), for different initial impurity momenta, \(p_I\) (see column labels). Here the postquench interaction is \(g_{BI}^f = 0.1 =g_{BB}\). The dashed lines indicate \(x_B = (\pm 1-\beta_f)c t\), with \(\beta_f\) the final velocity of the generated polaron provided as an inset label. (c), (d) The time-dependent overlap, \(| \langle \Psi_0 | \Psi(t) \rangle |\), of the post-quench many-body wavefunction, Eq. \eqref{Gross}, with the initial state, for varying \(g_{BI}\). The insets of (c), (d) provide the time-evolution of \(| \langle \Psi_0 | \Psi(t) \rangle |\) within a more extensive \(g_{BI}\) range. (e), (f) present the modulus and phase of the GA bath-wavefunction, \(\psi(x_B;t)\) respectively for \(g_{BI}^f = 0.1\), \(p_I = 0.4\) and \(t = 400\). For comparison (e) and (f) also provide the equilibrium profile of the polaron, Eq. \eqref{Hakim_solution}, with \(\beta = \beta_f = 0.64\). In all cases the system is confined in a ring of \(L =1600 n_0^{-1}\) and contains \(N_B = 1600\) while \(m_I = m_B\).}
\end{figure}

Typical spatiotemporal evolution patterns of the two-body correlation function, \(g^{(2)}_{IB} (0;x_B;t) = (L/n_{0}) \rho^{(2)}_{IB} (0;x_B;t)\), for an initial velocity of the impurity that does not exceed the speed of sound \(c = \sqrt{g_{BB} n_0 /m_r} \approx 0.45 \hbar n_0 / m_B\), are presented in Fig. \ref{fig:dynamicsFid} (a), (b).
The different panels correspond to varying initial impurity momenta, \(p_I\), but in all cases the  final interaction strength \(g^{f}_{BI} = g_{BB} = 0.1\) is kept fixed and we employ equal masses \(m_I = m_B\) for the impurity and the bath species.

In the case of a static impurity, \(p_I = 0\), it can be seen that the quench leads to initial (t < 10) emission of two \(\rho^{(2)}_{IB}(0;x_B;t)\) disturbances travelling away from the impurity with a velocity proximal to the speed of sound, \(\pm c\), see the dashed lines in Fig. \ref{fig:dynamicsFid} (a).
These disturbances subsequently break into structures, possessing an oscillatory two-body density pattern in space and being reminiscent of dispersive shock waves \cite{kamchatnov12_gener_disper_shock_waves_by,chang08_format_disper_shock_waves_by,leboeuf01_bose_einst_beams,hakim97_nonlin_schroed_flow_past_obstac_one_dimen}, see the box in Fig. \ref{fig:dynamicsFid} (a) and also Appendix \ref{sec:potential-appendix}.
In the vicinity of the impurity, \(r = 0\), a depletion of bath atoms emerges similarly to the case of a static polaron analyzed previously, see Sec. \ref{sec:static-polaron-GA} and \ref{sec:correlations}.
For later times, \(t \geq 300\), the two-body density \(\rho^{(2)}_{BI}(0;x_B;t)\) in the spatial extent of the impurity \(|x_{B}| < 30\), matches very well to the expected profile for a static, \(\beta =0\), polaron of Eq. \eqref{Hakim_solution}. 

This behavior can be explained due to the instantaneous quench and the sharpness of the \(\delta(r)\) interaction potential among the bath and the impurity.
In particular, it is well documented in the BEC literature \cite{scherer07_vortex_format_by_mergin_multip,denschlag00_gener_solit_by_phase_engin,becker08_oscil_inter_dark_dark_brigh} that rapidly switching on a potential within the spatial extent of a BEC leads to the phenomenon of phase imprinting\footnote{This means that due to the presence of the potential, the phase of the BEC shifts, leading to a flow of the bosonic density away from the repulsive potential. Note here that the amplitude of this phase disturbance increases with the decrease of the width of the perturbing potential. This effect is maximized for a \(\delta\)-shape as the one corresponding to the bath-impurity interactions.}.
The resulting disturbance in the vicinity of the impurity subsequently propagates outwards leading to the excitation of the bosonic host and the formation of dispersive shock wave structures. 
These excitations carry away the additional energy due to the quench allowing for the polaron to be formed behind them.
For more details on this mechanism see also Appendix \ref{sec:potential-appendix}.

For increasing impurity momentum, \(p_I = 0.4 \approx m_I c\), we observe a qualitatively different system response, see Fig. \ref{fig:dynamicsFid} (b).
Here, the two-body density disturbance emitted ``upstream'' (i.e. towards the direction of motion of the impurity) recedes from the impurity at a much slower pace than the corresponding ``downstream'' disturbance while the former has significantly larger amplitude than the latter.
These observations are explained in terms of the drag experienced by the moving impurity.
More specifically, it is known that if the velocity of a perturbing potential relative to a superfluid exceeds a certain critical value then the superfluidity of the environment is broken and the potential experiences a drag force.
The latter is analyzed in Ref. \cite{katsimiga18_many_body_dissip_flow_confin,khamis13_super_flow_bose_einst_conden,kamchatnov12_gener_disper_shock_waves_by} in the case that an external potential is dragged through a BEC.
Note that this external potential possesses a well-defined instantaneous position, independently of the exerted drag force.
However, the physical situation described here is slightly different because the impurity is a quantum particle that carries a definite kinetic energy.
Therefore, when a drag force emerges, it leads to the deceleration of the impurity up to the point that its velocity is so small that the drag force is nullified.
Except from the reduction of the impurity velocity, the drag force leads to Cherenkov-like\footnote{This term denotes the excitation of the BEC due to the locally supersonic motion of the impurity. This effect is analogous to the emission of electromagnetic radiation when electrons move through a dielectric medium with velocity greater than the phase velocity of light.} radiation \cite{susanto07_vceren_like_radiat_binar_super,carusotto06_bogol_vceren_radiat_bose_einst,el06_obliq_dark_solit_super_flow}. 
This leads to the amplification of the disturbances emitted ``upstream'' of the impurity and the emergence of an associated energy transfer process from the impurity to its bosonic environment.

The finite asymptotic velocity of the impurity, is indicated in \(\rho^{(2)}_{IB}(0;x_B;t)\) as a difference in the magnitude of the relative velocity of the emitted dispersive shock waves.
Indeed, by assuming that the disturbances travel with a velocity \(v_{DSW} \approx \pm c\) then in the impurity frame their velocities would be modified to \(v_{DSW} - v_p = (\pm 1 - \beta_f) c\), where \(\beta_{f} c\) is the velocity of the polaron. 
To estimate the final velocity of the polaron \(\propto \beta_{f}\), we fit the density and phase profile of the GA wavefunction \(\psi( | r | \leq 30;t)\), after an evolution time of \(t= 400\), to the corresponding analytic expression\footnote{Note here that the phase of \(\psi(r;t)\) is shifted so that \(\arg [\psi(r=0;t)] =0\) and the rest of the parameters in Eq. \eqref{Hakim_solution} are fixed to their corresponding values in the thermodynamic limit, namely \(n_0 = 1\) and \(\xi = (0.1)^{-1/2}\).}, Eq. \eqref{Hakim_solution}, for obtaining \(\beta_{f}\) and \(r_0\).
In Fig. \ref{fig:dynamicsFid} (b), we demonstrate that the above approximations are in exellent agreement with the motion of the dispersive shock waves.
Notice here, that the emitted structures realize a so-called ``light'' cone via which the correlations among the bath and the impurity are spread in the system after the quench \cite{lieb72_finit_group_veloc_quant_spin_system,cheneau12_light_cone_like_spread_correl}.
In particular, by examining the modulus (Fig. \ref{fig:dynamicsFid} (e)) and phase (Fig. \ref{fig:dynamicsFid} (f)) of \(\psi(r;t)\), we can verify that the corresponding profiles match the equilibrium polaron solution with \(\beta = \beta_f\), Eq. \eqref{Hakim_solution}, in the spatial extent between the shock waves.
Therefore, these excitations provide a means for transferring information regarding the generation of the Bose polaron throughout the BEC with a velocity equal to \(c\), see also the dashed lines in Fig. \ref{fig:dynamicsFid} (b).

\subsubsection{Time-dependent overlap: temporal orthogonality catastrophe}
\label{sec:orgf725e43}

Having appreciated, the main features of the two-body correlation dynamics, we now analyze their imprint on the time-dependent overlap \(| \langle \Psi_0 | \Psi(t) \rangle |\) for different \(g_{BI}^f\).
The quantity \(| \langle \Psi_0 | \Psi(t) \rangle |\) is commonly referred to as the fidelity between the \(| \Psi_0 \rangle\) and \(| \Psi(t) \rangle\) many-body states and it is related to the time-evolution of the quasi-particle residue, \(Z_{\rm p}\), of the polaron \cite{mistakidis20_many_body_quant_dynam_induc,mistakidis19_quenc_dynam_orthog_catas_bose_polar}.
By inspecting the static polaron case, \(p_I =0\), presented in Fig. \ref{fig:dynamicsFid} (c), we observe a very similar behavior as in the case of a parabolically trapped Bose-gas-impurity system examined in Ref. \cite{mistakidis19_quenc_dynam_orthog_catas_bose_polar}.
For interactions satisfying \(g_{BI} < g_{BB} = 0.1 \ll 2 \hbar c \approx 0.9\) the time-dependent overlap \(| \langle \Psi_0 | \Psi(t) \rangle |\) possesses a value proximal to \(1\), indicating that the state of the impurity after the quench is almost equivalent to the non-interacting one (Eq. \eqref{initial_state}).
Indeed, as it can be deduced from Fig. \ref{fig:comparisonMFMB} (d), (f), in this regime the residue of the polaron is \(Z_{\rm p} \approx 1\) and also its energy is proximal to \(E_{\rm p} \approx g_{BI} n_{0}\). 
These imply that the quench does not result in a pronounced production of excitations such as the dispersive shock waves exhibited in Fig. \ref{fig:dynamicsFid} (a), that would substantially affect \(| \langle \Psi_0| \Psi(t) \rangle |\) as we discuss below. 
For stronger interactions, \(| \langle \Psi_0 | \Psi(t) \rangle |\) changes drastically.
Around \(g_{BI} \approx g_{BB} = 0.1\), we find that \(| \langle \Psi_0 | \Psi(t) \rangle |\) is substantially depleted during the dynamics, reaching a finite value \(| \langle \Psi_0 | \Psi(t) \rangle | > 0\) for long times \(t > 300\), see Fig. \ref{fig:dynamicsFid} (c).
This behavior is inherently related to the emission of dispersive shock wave disturbances and the formation of the Bose polaron behind them as observed in Fig. \ref{fig:dynamicsFid} (a).

To explain this behaviour for intermediate interactions we have to examine the equilibrium properties of the Bose polaron and in particular its energy, Eq. \eqref{polaron_energy}. 
As shown in Fig. \ref{fig:analytics} (a), the non-linear correction terms in Eq. \eqref{polaron_energy} become important, leading to a sizable correction from the linear behavior observed for \(g_{BI} < g_{BB}\).
This leads to an energy surplus of the post-quench state, possessing \(E = g_{BI} n_0\), when compared to the corresponding polaronic state that the system eventually relaxes to.
Therefore, the emergence of the dispersive shock waves can be explained as a mechanism that carries the excess energy away from the region of the impurity.
The presence of these additional structures leads to the depletion of the time-dependent overlap \(| \langle \Psi_0 | \Psi (t) \rangle |\).
A similar behaviour occurs also for stronger interactions, \(g_{BI} > g_{BB} = 0.1\), where \(| \langle \Psi_0 | \Psi(t) \rangle |\) eventually saturates to zero.
Because of this the final state of the system is almost orthogonal to the initial one, therefore leading to the phenomenon of the temporal orthogonality catastrophe \cite{mistakidis21_radiof_spect_one_dimen_trapp_bose_polar,mistakidis20_pump_probe_spect_bose_polar,mistakidis20_many_body_quant_dynam_induc,mistakidis19_quenc_dynam_orthog_catas_bose_polar}.

Let us now comment on the influence of the initial impurity momentum on the time-dependent overlap \(| \langle \Psi_0 | \Psi(t) \rangle |\).
Figure \ref{fig:dynamicsFid} (d), depicts \(| \langle \Psi_0 | \Psi (t) \rangle |\) for a finite momentum impurity \(p_I = 0.4\), where a similar response to the static case, Fig. \ref{fig:dynamicsFid} (c), takes place.
The most important difference is observed for \(g_{BI} \approx g_{BB}\), where a larger suppression of \(| \langle \Psi_0 | \Psi(t) \rangle |\) occurs for \(p_I = 0.4\) than for \(p_I =0\).
The discrepancy of the moving impurity case when compared to the static one can be explained in terms of the additional drag force that emerges in the former scenario.
As already discussed above, the drag force leads to an impurity velocity smaller than the initial one since part of the initial momentum of the impurity is transferred to the ``upstream'' emitted dispersive shock wave excitation, see Fig. \ref{fig:dynamicsFid} (b).
This reduction of the impurity velocity during the dynamics leads to a further suppression of \(| \langle \Psi_0 | \Psi(t) \rangle |\) than the one observed for \(p_I = 0\), resulting in the appearance of the temporal orthogonality catastrophe phenomenon even in the case \(g_{BI} \approx g_{BB}\).

\subsubsection{Drag force and momentum transfer mechanism}
\label{sec:orgc7e9b47}

Let us now elaborate on the influence of the drag force in the time-evolution of the polaronic state. 
As already mentioned for a moving polaron the drag force reduces the velocity of the impurity up to a value where the drag force is nullified.
According to Ref. \cite{pavloff02_break_super_atom_laser_past_obstac,katsimiga18_many_body_dissip_flow_confin} the drag force can be approximated as
\begin{equation}
\label{drag-force}
F_D = \int \mathrm{d}r~ | \psi(r) |^2 \frac{\mathrm{d} V_{BI}}{\mathrm{d} r} = - g_{BI} \frac{\mathrm{d} | \psi (r) |^2}{\mathrm{d} r} \bigg|_{r = 0},
\end{equation}
where \(V_{BI}\) is the impurity potential perturbing the BEC, and it corresponds in our case to the bath-impurity interaction term, \(V_{BI} = g_{BI} \delta(r)\).
Equation \eqref{drag-force} indicates that the drag force is proportional to the derivative of the density in the vicinity of the impurity standing as a material barrier.
In the case of a polaron, Eq. \eqref{Hakim_solution} reveals that \(\frac{\mathrm{d} | \psi (r) |^2}{\mathrm{d} r} = 0\) for \(r=0\), and therefore the drag-force is zero.
This implies that a trivial upper bound that the final velocity of the impurity should satisfy is \(\beta_f < \beta_{\rm crit}\), in order to allow for polaronic solutions.
For impurities moving with subsonic velocities we can get a better upper bound for the final velocity of the polaron by considering the available values of the \(p_I\) for the final equilibrium polaronic state.
In the case that the polaronic state is created adiabatically, then the final state after the quench would possess a momentum \(p_I\) and the corresponding velocity \(\beta_{\rm p}(p_I)\), indicated in Fig. \ref{fig:ground_state} (a).
However, since we are considering an interaction quench this scenario is not realized and instead we would have a final momentum for the polaron \(p_I^f \leq p_I\) and a final velocity \(\beta_f \leq \beta_{\rm p}(p_I)\).

\begin{figure}[h]
\centerline{\includegraphics[width=1.0\textwidth]{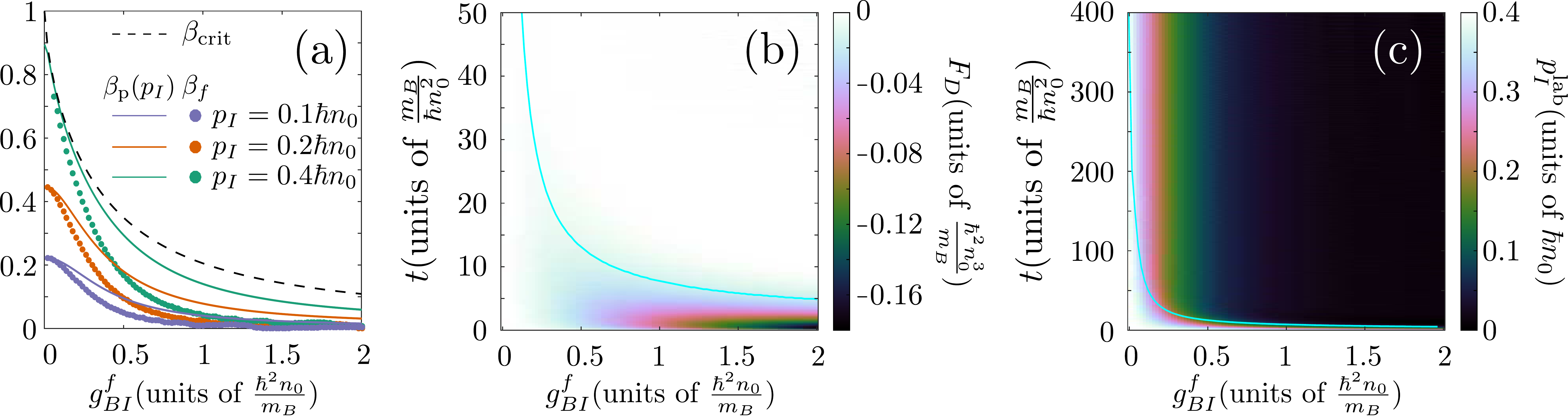}}
\caption{\label{fig:dynamics_velocities} Characterization of the drag force exerted on the impurity by the Bose gas. (a) The critical, \(\beta_{\rm crit}\) (dashed line) and equilibrium, \(\beta_{\rm p}(p_I)\) (solid lines) velocities of the Bose polaron with \(g_{BI} = g_{BI}^f\), compared to the final velocity of the polaron formed after the quench, \(\beta_f(p_I)\) (data points) for varying \(g_{BI}^f\). The parameters of the system are as in Fig. \ref{fig:dynamicsFid} and \(p_I\) is given in the legend. (b) Temporal evolution of the drag force exerted to an impurity, initially possessing \(p_I =0.4 \hbar n_0\), for different values of the post-quench interspecies interaction strength, \(g_{BI}^f\). (c) Time-evolution of the impurity momentum for \(p_I = 0.4 \hbar n_0\) and varying \(g_{BI}^f\). The solid lines in (b), (c) indicate the time that \(p_I^{\rm lab}\) becomes equal to the corresponding value for the equilibrium polaron with \(g_{BI} = g_{BI}^{f}\).}
\end{figure}

To justify the above let us clarify the role of the conserved quantity \(p_I\) in the dynamics.
According to Eq. \eqref{momentum-impurity-lab}, we have \(p_I = \langle \Psi(t) | \hat{p}_I^{\rm lab} | \Psi(t) \rangle + \langle \Psi(t) | \hat{p}_B | \Psi(t) \rangle\) and therefore only the sum of the impurity and bath momenta in the laboratory frame has to be conserved.
Recall that, the state of the system for long times corresponds to a polaron and two dispersive shock wave excitations that are far away from one another so that they do not interact, see Fig. \ref{fig:dynamicsFid} (b).
Due to the exerted drag force on the impurity and the consequent induced Cherenkov radiation, the upstream shock wave carries a larger (in magnitude) momentum than the downstream one.
Therefore, these two structures contribute a value \(\Delta p > 0\) to the total momentum, equal to the corresponding difference of their momenta.
This, in turn, implies that the momentum of the polaron for long times satisfies, \(p_I^f = p_I - \Delta p < p_I\) and due to the increasing tendency of \(\beta_{\rm p}(p_I)\) with \(p_I < m_I c\), see Fig. \ref{sec:moving-polaronGA} (a\textsubscript{2}), \(\beta_f \leq \beta_{\rm p}(p_I)\).   

The above arguments can be directly verified by our numerical calculations, see Fig. \ref{fig:dynamics_velocities} (a), where we compare the velocity after the quench, \(\beta_f\) (obtained by the same procedure as in Fig. \ref{fig:dynamicsFid} (a), (b)) to \(\beta_{\rm p}(p_I)\). 
This procedure yields that except for \(g_{BI} \sim 0\), \(\beta_f < \beta_p(p_I)\) holds independently of the value of \(p_I\), demonstrating the diabatic character of the polaron formation after an interaction quench of \(g_{BI}\).
Additionally, the dynamics becomes more diabatic as the postquench interaction strength is increased with the velocity of the polaron approaching a value of \(\beta_f = 0\) for strong \(g_{BI}\), independently of \(p_I\).
This more diabatic character of the dynamics with increasing \(g_{BI}\) can be understood by invoking Eq. \eqref{drag-force} implying that the amplitude of the drag force scales proportionally to the bath-impurity interaction strength.
According to the above, and as Fig. \ref{fig:dynamics_velocities} (b) testifies, the drag force is applied more abruptly to the impurity particle as \(g_{BI}\) increases, leading to a higher degree of excitation of the bath and hence larger momentum transfer, \(\Delta p_I\).
This momentum transfer can be directly probed in experiments by monitoring the momentum in the laboratory frame, \(p_I^{\rm lab}(t) = \langle \Psi (t) | \hat{p}_I^{\rm lab} | \Psi (t) \rangle\). 
Figure \ref{fig:dynamics_velocities} (c) indicates the decreasing tendency of \(p_I^{\rm lab} (t)\) with time for all interaction strengths.
Most importantly, even for small times \(t < 50\) and \(g_{BI} > g_{BB}\), \(p_{I}^{\rm lab} (t)\) becomes smaller than the corresponding equilibrium value for the polaron, see the corresponding lines in Fig. \ref{fig:dynamics_velocities} (b) and (c), demonstrating the existence of the momentum transfer mechanism.

\subsection{Dynamics of a supersonic impurity}
\label{sec:org8e6bcb4}

\begin{figure}[htbp]
\centerline{\includegraphics[width=1.0\textwidth]{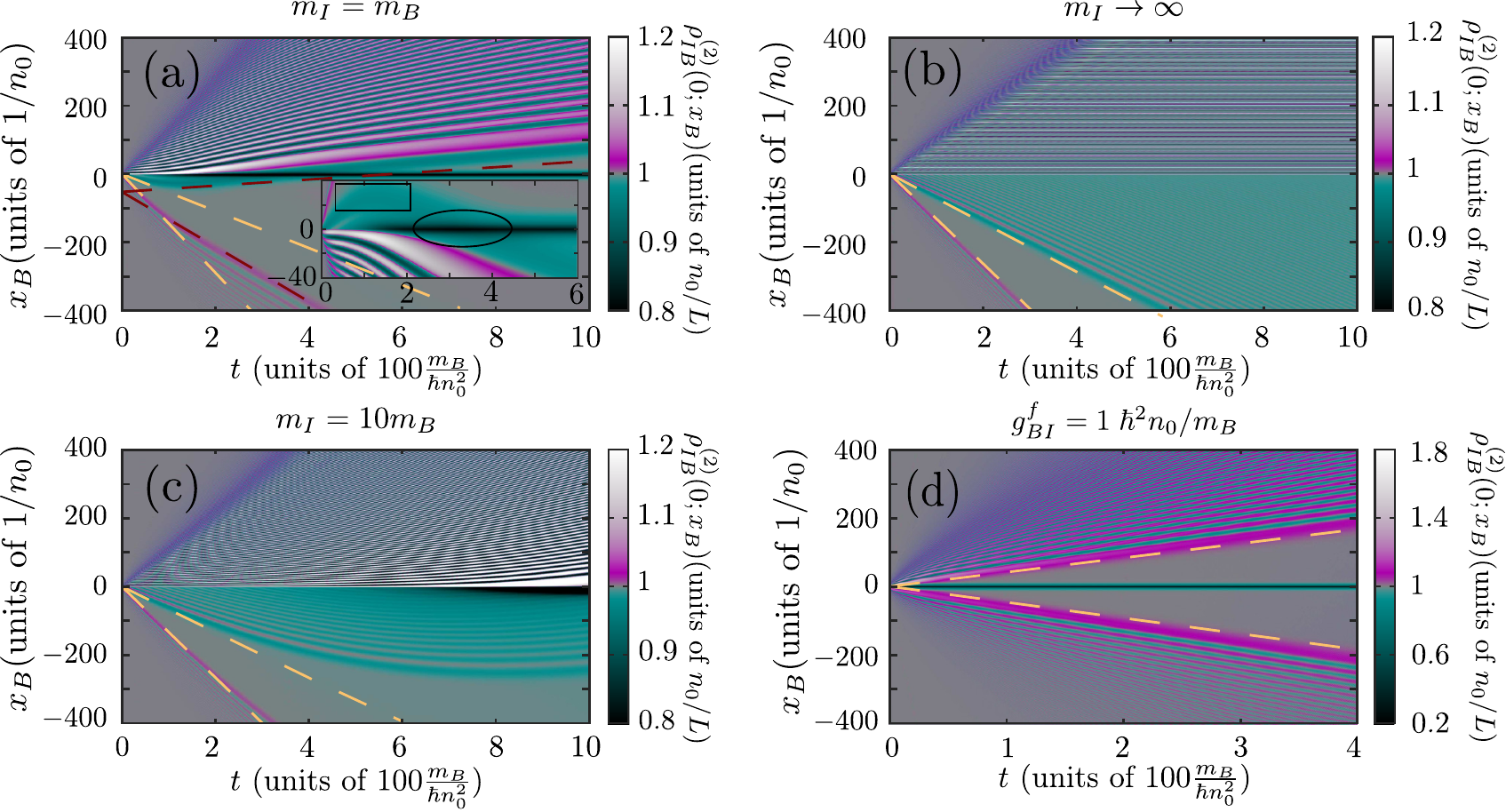}}
\caption{\label{fig:supersonic} Dynamics of an initially supersonically moving impurity. (a), (b), (c) Spatiotemporal evolution of the bath-impurity correlation function, \(\rho^{(2)}_{IB}(0;x_B)\), for a quench to \(g_{BI}^f = 0.07 \hbar^2 n_0 /m_B\) and \(p_I =1 \hbar n_0\). The inset of (a) provides a magnification of the corresponding bath-impurity correlation function in the vicinity of the impurity. The mass of the impurity is provided by the corresponding labels, while \(g_{BB} = 0.1 \hbar^2 n_0 /m_B\), \(N_B = 3200\) and \(L = 3200 n_0^{-1}\). The light dashed lines indicate \(x_I = (\pm c - p_{I}/m_I)t\) and the dark dashed lines in (a) correspond to \(x_I = (\pm 1 - \beta_f)c t +x_0\), with \(\beta_f = 0.7464\) the final velocity of the polaron and \(x_0 = 64 n_0^{-1}\) an offset selected for illustration purposes. (d) The time evolution of \(\rho^{(2)}_{IB}(0;x_B)\), for the same parameters as in (a) except for \(g_{BI}^f =1 \hbar^2 n_0/m_B\). The trajectories indicated by the dashed lines correspond to \(x_I = (\pm 1 - \beta_f)c t\), with \(\beta_f = 0.08\).}
\end{figure}

To conclude, we shall briefly comment on the case of a supersonically moving impurity.
Figure \ref{fig:supersonic} illustrates a characteristic example of the corresponding correlation dynamics, when \(p_I = 1\) and \(g_{BI}^f = 0.07 < g_{BB} = 0.1\).
We observe that for small times \(t < 100\) in addition to the emitted dispersive shock waves, also a density depletion takes place downstream of the impurity (see boxed area in the inset of Fig. \ref{fig:supersonic} (a)).
As the impurity slows down due to the exerted drag-force, the polaron starts to form at times \(t \approx 100 - 200\).
The depleted part of the density then collides at \(t \approx 400\) (see the encircled region in the inset of Fig. \ref{fig:supersonic} (a)) with the newly formed polaron lying at \(x_B =0\).
Eventually the density depletion overtakes the polaron ending up in the upstream region, \(x_B > 0\), for longer times, \(t > 600\) (see Fig. \ref{fig:supersonic} (a)).
Similarly to the subsonic cases the motion of the downstream shock wave (hardly visible in Fig. \ref{fig:supersonic} (a)) and the large amplitude density excitation emanating in the upstream region are moving with a velocity equal to the speed of sound.
Notice the agreement of the excitation trajectory with \(x_{\rm DSW} = (\pm 1 - \beta_{f}) c t + x_0\) for \(t > 300\), here \(\beta_{f} = 0.7464\) is the final velocity of the polaron found via fitting the polaron profile as discussed previously.

The origin of this additional excitation can be traced back to known properties for 1D BEC subjected to barrier dragging \cite{mistakidis19_dissip_correl_dynam_movin_impur}.
It is known  that for barrier velocities exceeding a threshold \(u_b > c\), stationary flow solutions (in the frame comoving with the barrier) exist, see the discussion in Ref. \cite{pavloff02_break_super_atom_laser_past_obstac,leboeuf01_bose_einst_beams}.
These solutions are characterized by a flat profile downstream of the barrier and a periodically modulating density in the upstream region.
However, for such structures in contrast to the polaron corresponding to Eq. \eqref{Hakim_solution}, the drag force exerted to the barrier is finite.
Consequently, since the impurity playing the role of the potential barrier possesses a finite momentum such flows cannot be stationary and have to decay when the velocity of the impurity becomes smaller than \(u_b\).
To substantiate the above claim, we solve the corresponding Gross-Pitaevskii equation in the frame of the potential barrier
\begin{equation}
    \label{GPE-std}
    \begin{split}
    i \hbar \frac{\partial}{\partial t} \psi(r) = \bigg[-\frac{\hbar^2}{2 m_B} \frac{\partial^2}{\partial r^2} +i v_0 \frac{\partial}{\partial r} +g_{BI} \delta(r) +g_{BB}(N_B-1) |\psi(r)|^2 \bigg]\psi(r),
    \end{split}
\end{equation}
where the constant velocity of the barrier, \(v_0\), is set to the initial velocity of the impurity.
In particular note that Eq. \eqref{GPE-std} is a reduction of Eq. \eqref{GPE} for \(m_{I} \to \infty\) and \(v_0 = p_{I}/m_I = \text{constant}\). 
Therefore, it corresponds to the asymptotic polaron solution for an infinitely heavy impurity, \(m_I \to \infty\).
Figure \ref{fig:supersonic} (b) presents the spatiotemporal evolution of the BEC density after a quench from \(g_{BI}=0\) to \(g_{BI}^f = 0.07\) and \(v_0 =1\), \(g_{BB} = 0.1\).
At the initial stages of the dynamics we observe the emission of the downstream dispersive shock wave (hardly visible in Fig. \ref{fig:supersonic} (b)).
However, the picture regarding the rest of the emerging structures is different than what was discussed for the \(m_I = m_{B}\) polaron.
In particular, we observe that upstream of the impurity a stationary oscillatory density pattern forms, reminiscent of the above mentioned solutions described in Ref \cite{pavloff02_break_super_atom_laser_past_obstac,leboeuf01_bose_einst_beams}.
Notice also that downstream of the impurity another dispersive shock wave structure is emitted.
The trajectories of the fronts of the two emitted shock waves indicate a relative velocity \(v_{\rm DSW} = (\pm c -v_{0})\) with respect to the impurity.
Therefore, these dispersive shock waves form a ``light'' cone, similarly to what we have observed for subsonically moving polarons.
However, since in this case \(v_0 > c \approx 0.32\), both shock waves lie in the downstream region of the barrier.

To relate the results of Eq. \eqref{GPE-std} to the case of a polaron, we consider a massive impurity with \(m_I = 10 m_B\).
The justification of this choice is that a massive impurity possesses larger inertia and therefore it is less susceptible to deceleration stemming from the drag-force exerted by the BEC.
This allows us to probe a possible intermediate mass regime for the cases depicted in Fig. \ref{fig:supersonic} (a) and (b).
Figure \ref{fig:supersonic} (c) shows the time evolution of \(\rho^{(2)}_{IB}(0;x_B)\) for a quench with the same parameters as in Fig. \ref{fig:supersonic} (a), (b).
For initial times, \(t < 200\), the two-body correlations exhibit the same structure as the one observed in the Gross-Pitaevskii case.
In particular notice the emission of the two downstream dispersive shock waves forming a ``light'' cone (\(v_{\rm DSW} = (\pm c - p_I/m_I)\)) and the quasi-stationary upstream oscillatory density pattern.
Subsequently, due to the finite momentum of the impurity and the exerted drag-force from the BEC the impurity slows down.
This can be verified by observing that for \(t > 200\) the position of the fronts of the emitted shock waves does not follow \(x_{\rm DSW} = (\pm c - p_I/m_I) t\).
Instead, they are shifted towards the impurity due to its reduced velocity.
Turning to long times, \(t> 600\), a sizable depletion of the BEC density in the vicinity of the impurity, \(x_B =0\) (see Fig. \ref{fig:supersonic} (c)), appears indicating the formation of the polaron.

By invoking the results of Fig. \ref{fig:supersonic} (b) and (c), we can interpret the initial stages of the dynamics, \(t < 100\), of Fig. \ref{fig:supersonic} (a) (referring to \(m_I = m_B\)) as the formation of a quasi-stationary supersonic BEC flow pattern and its decay when the velocity of the impurity becomes lower than \(u_{b}\).
After a transient time \(100 < t < 400\) where the polaron forms and slows down due to the drag-force it experiences, an equilibrium polaron state is reached for \(t > 400\) where the drag-force is nullified, similarly to the case of an initially subsonic impurity.

Finally we comment that the dynamics of the system exhibits a similar behavior as the one observed in Fig. \ref{fig:supersonic} (a), as long as, the interspecies interaction strength, \(g_{BI}\), is sufficiently weak.
Indeed, it is known \cite{pavloff02_break_super_atom_laser_past_obstac,leboeuf01_bose_einst_beams} that for larger interspecies interactions (or equivalently the barrier heights) the velocity threshold for the formation of stationary supersonic flow, \(u_b\), increases.
In addition, the amplitude of the drag-force, Eq. \eqref{drag-force}, is proportional to \(g_{BI}\), yielding a rapid deceleration of the impurity for large interspecies interactions.
Accordingly, for strong bath-impurity repulsions no stationary supersonic flow can be approached during the dynamics, since both the threshold \(u_b\) increases and the deceleration of the impurity becomes more prominent.
In this case, the dynamics of supersonically moving impurities is qualitatively similar to the regime \(p_I \approx m_I c\), compare Fig. \ref{fig:supersonic} (d) to Fig. \ref{fig:dynamicsFid} (b).

\section{Conclusions \label{sec:conclusions}}
\label{sec:org9a65b9f}

We have examined the stationary and dynamical properties of the 1D Bose polaron in the absence of external confinement.
It has been argued that, the stationary properties of the Bose polaron can be reliably evaluated within the GA approach for the case of a weakly interacting Bose gas \cite{jager20_stron_coupl_bose_polar_one_dimen,will21_polar_inter_bipol_one_dimen,panochko19_mean_field_const_spect_one,volosniev17_analy_approac_to_bose_polar}.
Within this approximation all non-trivial bath-bath correlations are neglected and the bath-impurity two-body correlations are variationally optimized.
By comparing with the correlated MCTDHB approach, we verify that the GA adequately captures important properties of the polaron such as its energy, effective mass and the bath-impurity two-body correlation profiles.
However, it is found that the residue is overestimated within GA as it neglects the quantum depletion of the BEC background.
Importantly, regarding a moving impurity, we have demonstrated that the character of the equilibrium many-body state crossovers from a polaronic quasi-particle to a collective excitation, having the form of a dark-bright soliton.
Indeed, for small interactions and momenta a polaron is generated and characterized by a localized depletion of the two-body bath-impurity correlation when the corresponding particles are in close proximity. 
In the opposite case of strong interactions or large momenta, the state of the mixture is similar to a stationary dark-bright soliton.

Regarding the dynamical response of the system we show that the phenomenon of the temporal orthogonality catastrophe which has been originally observed in confined polaron systems \cite{mistakidis21_radiof_spect_one_dimen_trapp_bose_polar,mistakidis20_pump_probe_spect_bose_polar,mistakidis20_many_body_quant_dynam_induc,mistakidis19_quenc_dynam_orthog_catas_bose_polar} generalizes to the homogeneous case (see also \cite{jager21_stoch_field_approac_to_quenc}).
In all cases, the system approaches an equilibrium polaron state in the long time dynamics accompanied by additional excitations induced by the quench.
In particular, for a static impurity the many-body wavefunction of the system becomes orthogonal to the corresponding non-interacting one for long timescales, despite the fact that the corresponding polaron state possesses a finite overlap to the non-interacting one \cite{volosniev17_analy_approac_to_bose_polar}.
For moving impurities the temporal orthogonality catastrophe is more pronounced since the drag force leads to the deceleration of the impurity.
Dispersive shock wave structures play an important role in the quenched polaron dynamics as they provide the means to transfer the excess energy due to the quench away from the spatial extent of the impurity allowing for the eventual relaxation of the system to an equilibrium polaron configuration.
Even in the case of a supersonically moving impurity a final equilibrium polaron configuration is reached.
However, the timescale needed for the slow-down of the impurity depends crucially on its mass.
The emission of these non-linear structures in the time-evolution highlights the importance of non-linear and non-perturbative processes for understanding the dynamics of impurity systems.
In addition to the above, the generality of the temporal orthogonality catastrophe mechanism for abrupt interaction quenches mandates a different experimental protocol relying on adiabatic transfer to the polaron configuration \cite{mistakidis21_radiof_spect_one_dimen_trapp_bose_polar} for realizing strongly interacting Bose polaron states.

There are several avenues for further research that can be pursued in future studies.
In particular, all the results presented herein refer to the weak interaction regime of the Bose gas where its state can be well approximated as a BEC and its excitations treated within the Bogoliubov approximation.
For stronger bath-bath interactions, where the elementary excitations of the Bose gas do not follow the Bogoliubov approximation \cite{lieb63_exact_analy_inter_bose_gas_II}, it is intriguing to examine the applicability of the GA approach and its limitations.
In addition, in this interaction range particle-hole excitations, namely the type II excitations of the Lieb-Liniger model \cite{lieb63_exact_analy_inter_bose_gas_I}, become significant and it is therefore interesting to inspect whether they contribute to the modification of the polaronic quasiparticle or the emergence of a distinct type of excitation.
Our findings indicating the importance of non-linear dynamics for characterizing the fate of the polaron might also be important for the case of higher dimensions, where recent studies indicate the possibility of the temporal orthogonality catastrophe \cite{guenther21_mobil_impur_bose_einst_conden_orthog_catas,drescher20_theor_reson_inter_impur_bose_einst_conden}.
This is particularly important since phenomena related to pattern formation and the emergence of drag force are well known for two- and three-dimensional systems \cite{PitaevskiiStringari2016}.
For instance, the creation and dynamics of structures such as oblique solitons and vortices \cite{el06_obliq_dark_solit_super_flow} might be relevant for understanding the quench induced dynamics of the Bose polaron in two dimensions and the temporal orthogonality catastrophe in such systems.

\begin{acknowledgements}
SIM gratefully acknowledges financial support in the framework of the Lenz-Ising Award of the University
of Hamburg.
 This work is funded by the Cluster of Excellence
‘Advanced Imaging of Matter’ of the Deutsche Forschungsgemeinschaft (DFG) - EXC 2056 - project ID
390715994. 
\end{acknowledgements}
\appendix

\section{Bosonic momentum renormalization \label{sec:renorm}}
\label{sec:org205d51f}

\begin{figure}[h]
    \centering
    \includegraphics[width=1.0\textwidth]{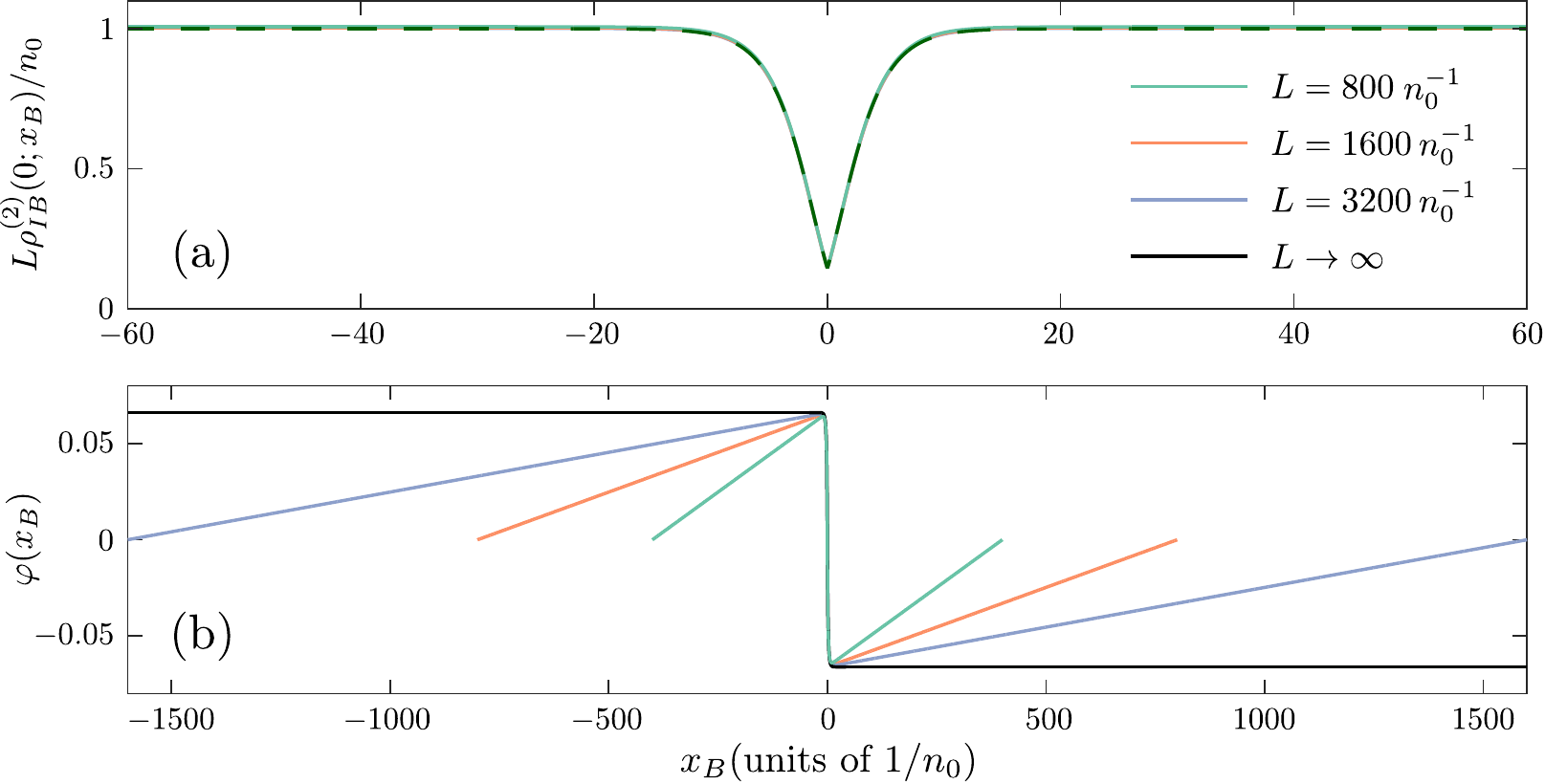}
\caption{\label{fig:phase_shift} Convergence of the GA Bose polaron solution to the \(N,L \to \infty\) limit. (a) Bath-impurity correlation function, \(\rho^{(2)}_{IB}(0;x_B)\) for \(g_{BI} = 1 \hbar^{2} n_0^{-1}/m_B\), \(g_{BB} = 0.1 \hbar^2 n_0/m_B\), \(p_I = 0.1 \hbar n_0\) and \(m_I = m_B\) but different ring lengths \(L\) (see legend). In order to keep \(n_0 =1\) in our calculations, we demand \(N_B = n_0 L\), while the spatial region \(|x_B| > 60 n_0^{-1}\) is not depicted, since  \(\rho^{(2)}_{IB}(0;x_B) \approx \rho^{(2)}_{IB}(0;60)\). (b) The phase profile, \(\phi(x_B) = \arg(\psi(x_B))\), of the solutions for the above mentioned parameters and for the same varying values of \(L\).}
\end{figure}

To appreciate the physical context of the unconventional boundary conditions of Eq. \eqref{Hakim_solution} we consider a system confined in an 1D ring of perimeter \(L \gg \xi\).
In this case the phase of \(\psi_{L}(r)\) satisfies \(\varphi_{L}(r)=\varphi_{L}(r+L)\) and therefore the solution of Eq. \eqref{Hakim_solution} cannot be embedded in this finite system.
However, the system for \(|r| \lessapprox \xi\), should behave in a similar manner to Eq. \eqref{Hakim_solution}, since the boundary conditions should not alter the behavior of the system at this spatial scale.
In particular, the convergence of \(\rho^{(2)}_{IB}(0;x_B)\) to the limit \(L \to \infty\) is observed already for \(L = 800\), see Fig. \ref{fig:phase_shift} (a).
This implies that also in this setting a phase shift occurs.
Indeed, such phase-shifts in the vicinity of the impurity can be observed for \(r \approx 0\) in the numerical solution of Eq. \eqref{GPE} for a system confined in a ring of finite perimeter, see Fig. \ref{fig:phase_shift} (b).

To compensate for this phase shift a phase gradient appears in the ring solution at \(r \gg \xi\) so that \(\varphi_{L}(\pm L/2)=0\).
This effect is captured in Fig. \ref{fig:phase_shift} (b) by examining different values of \(L\).
In addition, it can be clearly seen that this gradient decreases for increasing \(L\) and therefore in the case that \(L \to \infty\) the slope of this gradient is nullified.
Importantly, this alteration of the phase profile in the case of finite systems, results in a contribution to the bath momentum since the latter is defined as
\begin{equation}
    \label{<+label+>}
    p_B=-i \hbar N_B \int \mathrm{d} r~\psi^*(r) \frac{\mathrm{d}\psi(r)}{\mathrm{d}r} \approx \hbar \int \mathrm{d}r~ n(r) \frac{\mathrm{d}\varphi(r)}{\mathrm{d}r},
\end{equation}
where we have employed that \(\psi(r) = \sqrt{n(r)/N_B} e^{i \varphi(r)}\) and assumed that \(\frac{\mathrm{d} n(r)}{\mathrm{d} r} =0\) in the spatial extent where the counterflow occurs, \(r \gg \xi\).
Since the phase gradient occurs for large \(r \gg \xi\), where \(n(r) \to n_0\) and \(\frac{\mathrm{d} \varphi(r)}{\mathrm{d} r} \approx \text{constant}\), its momentum contribution is finite and characteristic of the phase difference created by the solution \(\Delta \varphi = \varphi(r = L/2) - \varphi(r = -L/2)\).
In particular, as evident in Fig. \ref{fig:phase_shift} (b) we can approximate \(\frac{\mathrm{d} \varphi(r)}{\mathrm{d} r} = - \frac{\Delta \varphi}{L}\).
The above imply that the momentum of the
system in the thermodynamic limit \(L \to \infty\) is shifted by a finite amount from the result
obtained by integrating the wavefunction of Eq. \eqref{Hakim_solution} such that
\begin{equation}
    \label{<+label+>}
    p_B=- i \hbar N_B \int_{-\infty}^{+\infty} \mathrm{d} r~\psi^*(r) \frac{\mathrm{d}\psi(r)}{\mathrm{d}r} - \hbar n_0 \Delta\varphi.
\end{equation}
The term \(p_{b.c.}=-\hbar n_0 \Delta \varphi\) is a characteristic shift caused by the unconventional
boundary conditions of the solution, Eq. \eqref{Hakim_solution}, and should always be added to the
``bare'' part stemming from the integration of the corresponding wavefunction.
This ``renormalization'' of the bosonic momentum is well-known in the literature, for a more
detailed discussion we refer the interested reader to Ref. \cite{PitaevskiiStringari2016}.

\section{The impact of the interspecies interaction potential \label{sec:potential-appendix}}
\label{sec:org8df63b6}

As discussed in the main text, dispersive shock wave excitations are emitted from the spatial regime of the impurity following an interspecies interaction quench.
Here we will elaborate on the origin of such excitations by comparing the case of a zero-range \(\delta\)-shaped interaction potential with finite width ones.
The key to understand the emission of non-linear patterns at initial times is the concept of phase imprinting and the relation of the phase of the BEC with its flow.
It is well known \cite{mistakidis21_radiof_spect_one_dimen_trapp_bose_polar,GrusdtDemler2015} that the typical time-scale for the formation of excitations in a BEC is of the order of \(\xi / c \sim \hbar / \mu\), where \(\xi\), \(c\) and \(\mu\) are the healing length, speed of sound and chemical potential of the BEC respectively.
Therefore, the bath atoms cannot react to any change of the system parameters occurring much faster than this time-scale.

Similarly, to Ref. \cite{mistakidis21_radiof_spect_one_dimen_trapp_bose_polar} this allows us, for \(t \ll \hbar/\mu\), to neglect the effect of terms proportional to \(\frac{\partial}{\partial r_k}\) appearing in the Hamiltonian \(\hat{H}_{\rm LLP}\), Eq. \eqref{hamilt_LLP}.
Within the GA the above imply that the equation of motion of Eq. \eqref{GPE} reduces to
\begin{equation}
\label{eq:1}
i \hbar \frac{\partial}{\partial t} \psi(r;t) = \bigg[ V_{BI}(r) + \underbrace{g_{BB} | \psi(r;t) |^2}_{\approx \mu} - \mu  \bigg] \psi(r;t),
\end{equation}
which can be solved yielding a time-evolution \(\psi(r;t) = e^{- \frac{i}{\hbar} V_{BI}(r) t} \psi(r;0)\) for the variational single-particle wavefuction of the bath.
In this solution, a spatially dependent shift of the BEC phase proportional to the local value of the interaction potential appears.
The phase of the BEC wavefunction, and hence the above mentioned shift, is important because it dictates the local velocity of the BEC flow according to 
\begin{equation}
\label{superflow}
v_{\rm superflow}(r;t) \equiv \frac{\hbar}{m_B} \frac{\partial}{\partial r} \varphi(r;t) \overset{t \ll \hbar/\mu}{=} \frac{1}{m_r} \left( - \frac{\partial V_{BI}(r)}{\partial r} \right) t,
\end{equation}
where \(\varphi(r;t) = \arg( \psi(r;t))\).
Equation \eqref{superflow} indicates that steep interaction potentials, lead to large values for the flow velocity of the BEC and importantly also large gradients of the flow velocity.
When this relative flow of the BEC becomes comparable to the speed of sound \(c\) then the superfluidity of the medium is broken and additional excitations emerge, according to the Landau criterion for superfluidity.
This procedure for inducing non-linear excitations, commonly referred to as phase imprinting, has been widely employed experimentally for the creation of non-linear structures such as vortices and solitons \cite{scherer07_vortex_format_by_mergin_multip,denschlag00_gener_solit_by_phase_engin,becker08_oscil_inter_dark_dark_brigh}.
Importantly for our discussion, the generation of dispersive shock waves by rapidly switching on a repulsive interaction potential has been demonstrated in Ref. \cite{chang08_format_disper_shock_waves_by}.
In that work the steepness of the perturbing potential was controlled by keeping its width fixed and increasing its amplitude. 
Below we will take a complementary approach where the steepness is increased by keeping \(g_{BI} = \int \mathrm{d} r~ V_{BI}(r)\) fixed, while reducing the overall width of the interspecies interaction potential.
More specifically, we compare \(V_{BI}(r) = g_{BI} \delta(r)\) with the Gaussian-shaped potential 
\begin{equation}
\label{Gaussian_potential}
V_{BI}(x) = \frac{g_{BI}}{\sqrt{2 \pi} w} e^{- \frac{x^2}{2 w^2}},
\end{equation}
where \(w\) parametrizes the finite width of the interaction.
Note here that also \eqref{Gaussian_potential} reduces to the \(\delta\)-function limit for \(w \to 0\).
In this context our discussion below outlines the extrapolation of the concepts developed in \cite{chang08_format_disper_shock_waves_by} to the case of potentials with an infinitesimal range.

\begin{figure}[htbp]
\centerline{\includegraphics[width=1.0\textwidth]{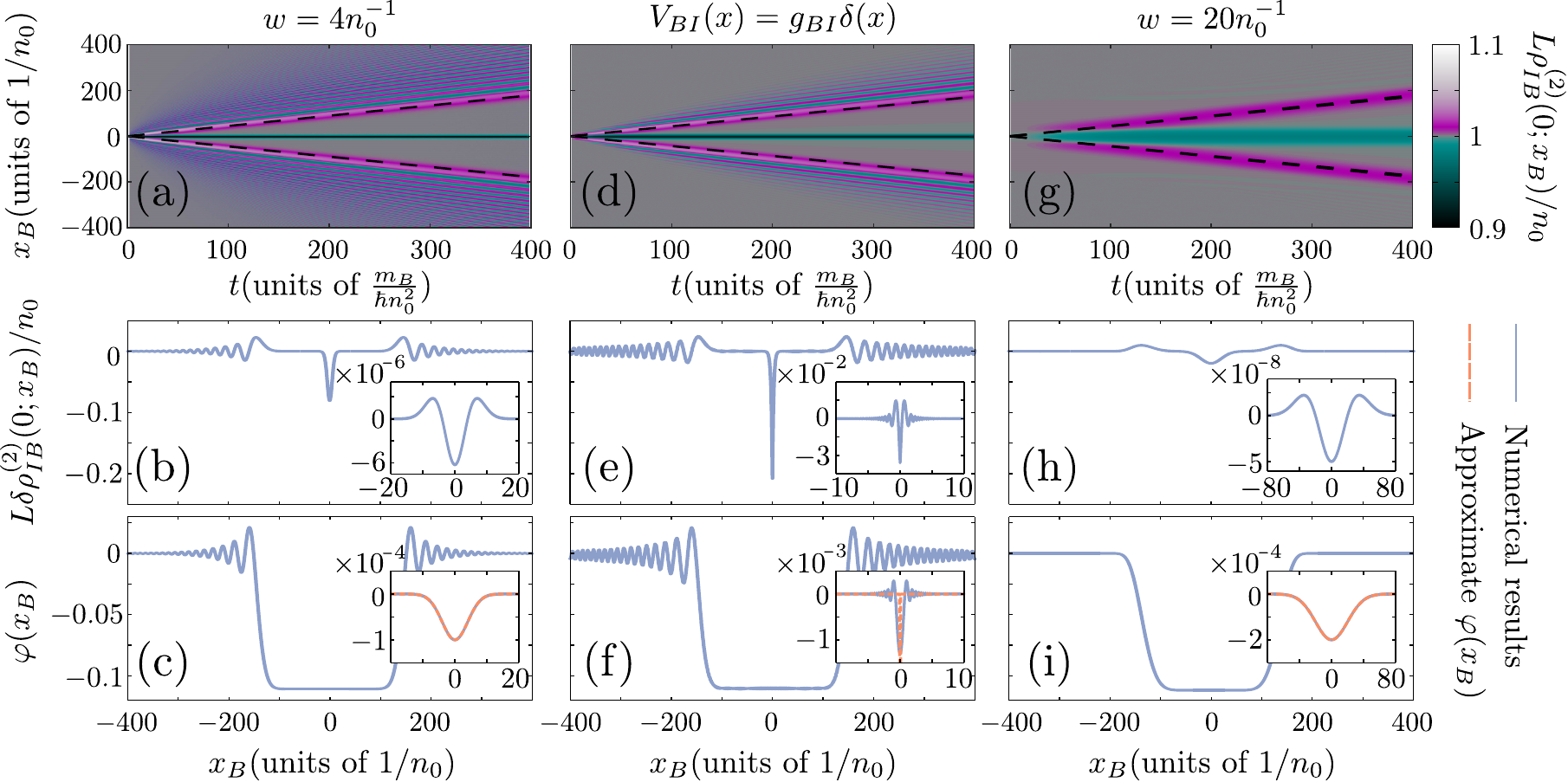}}
\caption{\label{fig:widths} Emergence of dispersive shock waves for short-range interaction potentials. (a) Spatiotemporal evolution of \(\rho^{(2)}_{IB}(0;x_B)\) for \(g_{BI}^f = 0.1 \hbar^2 n_0/m_B = g_{BB}\), \(p_I = 0\), \(L = 1600 n_0^{-1}\) and \(N_B = 1600\) in the case that a Gaussian bath-impurity interaction potential with width \(w =4 n_0^{-1}\) is employed. (b) Same as (a) but for the particular time instant \(t =300 m_B/(\hbar n_0^2)\).  (c) the phase profile corresponding to (b). The insets of (b), (c) correspond to \(t = 0.1 m_B/(\hbar n_0^2)\). (d), (e) and (f) correspond to the same quantities as in (a), (b) and (c) respectively, calculated for the same set of parameters. However, here the bath-impurity potential corresponds to \(\delta\)-function. (g), (h), (i) the same quantities as in (a), (b), (c) but for a wider Gaussian potential \(w =20 n_0^{-1}\). The insets of (c), (f) and (i) in addition to the GA numerical results also indicate the approximate profile expected from the phase imprinting of the impurity potential (see text).}
\end{figure}

Figure \ref{fig:widths} compares the polaron formation dynamics after a quench of the bath-impurity interaction strength to \(g_{BI}^f = 0.1 = g_{BB}\) for \(N_B = 1600\) bath atoms in a ring with \(L = 1600 n_0^{-1}\) for the different interaction potentials. 
Here, an initially static impurity, \(p_I = 0\), is considered possessing \(m_I = m_B\).
Figure \ref{fig:widths} (a) presents the time-evolution of \(\rho^{(2)}_{BI}(0;x_B)\) for a Gaussian interaction potential, Eq. \eqref{Gaussian_potential}, with width \(w = 4 \gtrapprox \xi \approx 3.16\).
In particular, Fig. \ref{fig:widths} (a) demonstrates the emission of dispersive shock waves, which are moving with velocity proximal to \(c\) (see the dashed lines in Fig. \ref{fig:widths} (a)). 
The oscillatory density pattern associated with these structures can be clearly seen in Fig. \ref{fig:widths} (b) for \(t = 300\) and \(|x_B | > 100\). 
It is associated with a corresponding oscillatory phase as demonstrated in Fig. \ref{fig:widths} (c).
These results are in agreement with previous studies on dispersive shock wave patterns \cite{hakim97_nonlin_schroed_flow_past_obstac_one_dimen,kamchatnov12_gener_disper_shock_waves_by}.
The behavior of the system for short times, \(t = 0.1 \ll \xi/c \approx 7.07\) elucidates the mechanism for the generation of these structures, see the insets of Fig. \ref{fig:widths} (b) and (c).
In particular, the expected profile 
\begin{equation}
\label{approx-phas-prof}
\varphi(x_B,t) = - \frac{t}{\hbar} V_{BI}(x).
\end{equation}
describes well the phase of the system, see the inset of Fig. \ref{fig:widths} (c).
Also as a consequence of Eq. \eqref{superflow}, we can observe that even at such short times a small portion of the BEC atoms (notice the \(10^{-6}\) scale in the inset of Fig. \ref{fig:widths} (b)) have already moved away from the impurity following the gradient of the phase.
The above outcomes are inline with the experiment of Ref. \cite{chang08_format_disper_shock_waves_by}.
Indeed, due to the phase imprinting at initial times, the bath particles are forced to move away from the impurity building up the wavepackets seen at \(x_B \approx 5\) in the inset of Fig. \ref{fig:widths} (b). 
Since the accumulated phase, Eq. \eqref{approx-phas-prof}, increases linearly with time the velocity of the flow forming these wavepackets increases and can thus reach supersonic speeds, when compared to the stationary flow away from the impurity, leading to the formation of dispersive shock waves.
As the steepness of the potential increases by reducing the value of \(w\) this phenomenon is amplified due to the faster increase of the superflow velocity in the spatial extent of the impurity, and therefore in the \(\delta\)-potential limit it becomes maximal motivating the occurrence of dispersive shock wave structures, also in this case.

To verify the above expectation we next focus to the case of the \(\delta\)-potential. 
The phenomenology in that case is similar to the Gaussian-potential one (see Fig. \ref{fig:widths} (d)), which is also supported by examining the density, Fig. \ref{fig:widths} (e), and phase, Fig. \ref{fig:widths} (f), profiles e.g. for \(t = 300\).
In the \(\delta\)-potential case, we cannot, however, find evidence for the mechanism of phase imprinting since for \(t = 0.1 \ll \xi/c \approx 7.07\).
Indeed, the density of the BEC in the vicinity of the impurity is significantly disturbed (see the inset of Fig. \ref{fig:widths} (e)) while the corresponding phase does not match the expected profile of Eq. \eqref{approx-phas-prof}.
To explain these apparent discrepancies one has to consider the \(\delta\)-potential as the asymptotic limit of a progression of \(V_{BI}(x)\) characterized by reducing width.
Indeed, as the width of the potential, \(w\), decreases, phase imprinting and the consequent generation of dispersive shock waves occur for smaller times.
In the asymptotic case of a \(\delta\) potential these processes are exhibited within an extremely small timescale resulting in the signatures of the dispersive shock waves already appearing for \(t = 0.1\), compare Fig. \ref{fig:widths} (e) and \ref{fig:widths} (f) with their corresponding insets. 

To make a more explicit connection to the experimental results of Ref. \cite{chang08_format_disper_shock_waves_by}, let us briefly comment on the case of a potential with very large width, \(w = 20 \gg \xi\).
In this case the behavior of the system is qualitatively different than for the previous cases, see Fig. \ref{fig:widths} (g), as no dispersive shock waves are produced.
Indeed, the emitted excitations refer to density modulations of the BEC within a length scale much larger than \(\xi\).
Because of this and due to the linearity of the Bogoliubov dispersion relation for quasimomenta \(k < 1/\xi\), we can conclude that such excitations are a superposition of sound waves, propagating away from the impurity with a group velocity given by the speed of sound \cite{joseph07_measur_sound_veloc_fermi_gas,andrews97_propag_sound_bose_einst_conden}.
The smoothness of the corresponding density profile can be verified by observing the density and the phase of the BEC at \(t=300\), see Fig. \ref{fig:widths} (h) and (i) respectively.
At first glance, the reason behind the generation of these excitations is not evident since for initial times the structures emerging in the density and phase of the BEC are qualitatively equivalent to the case of \(w = 4\), compare the insets of Fig. \ref{fig:widths} (b) and (c) with Fig. \ref{fig:widths} (h) and (i) respectively.
Here the significant quantitative difference in the amplitude of these patterns is the cause for the qualitatively different long time behavior that these two systems exhibit.
In particular, for \(w =20\), the variation of the phase is much smoother and it increases much slower in time than in the case of \(w = 4\).
This allows the bath particles to travel away from the impurity at much smaller speeds than in the case of a narrower potential and therefore the velocity of the superflow can smoothly decay to zero as we get away from the impurity.
Consequently, no non-linear excitations are produced since the Landau criterion is never violated and the only structure that gets emitted is a small density disturbance corresponding to the above mentioned sound waves.

\section{Details on the computational techniques \label{sec:mlx}}
\label{sec:org1a431bf}

To examine the non-equilibrium dynamics of the system we numerically solve the GA equation of motion, Eq. \eqref{GPE}, for a finite number of particles so that \(n_0 = N_B/L = 1\).
Recall that in this case the chemical potential, \(\mu(t)\), corresponds to a Lagrange multiplier, that has to be evaluated by demanding that the particle number is conserved.
The process outlined above, allows us to cast the equation-of-motion in the single-particle Schrödinger type equation
\begin{equation}
\label{pseudo-Schroedinger}
i \frac{\partial}{\partial t} | \psi (t) \rangle = \left( \hat{\mathcal{H}}[\psi(r;t)] - \langle \psi(t) | \hat{\mathcal{H}}[\psi(r;t)] |\psi(t)  \rangle \right) | \psi(r;t) \rangle,
\end{equation}
with \(\hat{\mathcal{H}}[ \psi(r;t)] =- \frac{\hbar^2}{2 m_B} \frac{\partial^2}{\partial r^2} + \frac{i \hbar^2 k_0(t)}{m_I} \frac{\partial}{\partial r} + g_{BI} \delta(r) + g_{BB} (N_B -1) | \psi(r;t) |^2\), denoting the effective single-particle Hamiltonian and \(\psi(r) \equiv \langle r | \psi(t) \rangle\).
Then the effective Schrödinger equation of Eq. \eqref{pseudo-Schroedinger} is discretized by employing an exponential discrete variable representation \cite{littlejohn02_gener_framew_discr_variab_repres_basis_sets}.
For the corresponding time-evolution we use the standard fourth order Runge-Kutta integrator.
Notice here, that the employed basis set intrinsically introduces periodic boundaries at both ends of the potential.
An advantage of the exponential discrete variable representation is that the first and second derivative matrices of the corresponding basis refer to the Fourier ones, allowing us to employ the Fast-Fourier-Transform algorithm for numerical efficiency.
In place of the \(\delta\)-potential an approximation of it is employed, namely
\begin{equation}
\label{kronecker_potential}
\left[ V_{BI} \right]_i = \frac{g_{BI}}{\Delta x} \delta_{i, \frac{n}{2}},
\end{equation}
where \(\delta_{ij}\) is the Kronecker delta, \(i= 0,2,...,n-1\) represents the index of each of the \(n\) grid points located at \(x_i = -\frac{L}{2} + \frac{i L}{n}\), and \(\Delta x = L/n\) is the grid spacing.
Within the discrete variable representation framework it can be shown that the approximation for the \(\delta\)-potential of Eq. \eqref{kronecker_potential} is the variationally optimal.
To estimate the validity of our numerical results, we repeat the calculations for different spatial, \(\Delta x\), and temporal, \(\Delta t\), discretizations.
We have verified that \(\psi(r;t)\) for time-intervals \(t < 400\) becomes independent of the discretization for \(\Delta t = 0.0005\) and \(\Delta x = 25/256\), where dimensionless units \(\hbar = m_B = n_0 =1\) are employed.

To estimate the impact of correlations in the ground state properties of the Bose polaron we utilize the 
Multi-Layer Multi-Configuration Time-Dependent Hartree method for atomic mixtures (ML-MCTDHX) \cite{cao17_unified_ab_initio_approac_to}. 
The key idea of ML-MCTDHX lies on the usage of a time-dependent and variationally optimized 
many-body basis set, which allows for the optimal truncation of the total Hilbert space. 
Since here we simulate only the Bose gas part of the many-body wavefunction the ML-MCTDHX method,
reduces to the simpler Multi-Configuration Time-Dependent Hartree method for bosons (MCTDHB) approach \cite{alon08_multic_time_depen_hartr_method_boson,alon07_unified_view_multic_time_propag}.
Within the latter the ansatz for the bath many-body wavefunction, \(| \Psi_B(t) \rangle\), is taken as a linear combination of time-dependent permanents \(|n_1,n_2, \dots, n_M (t) \rangle\),
\begin{equation}
\label{eq:3}
| \Psi_B (t) \rangle = \sum_{n_1, n_2, \dots, n_M | \sum_{k =1}^M n_k = N_B} A_{\vec{n}}(t) | n_1, n_2, \dots, n_M (t) \rangle,
\end{equation}
with time-dependent weights \(A_{\vec{n}}(t)\). 
In turn, each time-dependent permanent is expanded in terms of \(M\) time-dependent variationally optimized single-particle functions \(\phi_k (r;t)\) as follows
\begin{equation}
\label{eq:2}
\begin{split}
\langle r_1,&\dots,r_{N_B} | n_1, n_2, \dots, n_M (t) \rangle =\\
 &|\left( \prod_{k=1}^{M}n_k! \right)^{-\frac{1}{2}} \sum_{i=1}^{N_B!} \prod_{j=1}^{n_1} \phi_1(r_{\mathcal{P}_i(j)};t) \prod_{j=n_1+1}^{n_1+n_2} \phi_2(r_{\mathcal{P}_i(j)};t) \dots \prod_{j=1+\sum_{k=1}^{M-1} n_k}^{N_B} \phi_M(r_{\mathcal{P}_i(j)};t),
\end{split}
\end{equation}
where \(\mathcal{P}_i\) is the operator performing the \(i\)th permutation of \(\{ 1, 2, \dots, N_B \}\).
For our numerical implementation the single-particle functions are expanded within a primitive basis corresponding to the exponential discrete variable representation that we also use in the GA case.
The time-evolution of the \(N_B\)-body wavefunction under the effect of the Hamiltonian \(\hat{H}_{\rm LLP}\) reduces to the determination of the \(A\)-vector coefficients and the single-particle functions, which follow the variationally obtained equations of motion \cite{alon08_multic_time_depen_hartr_method_boson,alon07_unified_view_multic_time_propag}. 
Let us note here that in the limiting case of \(M=1\), the method reduces to Eq. \eqref{pseudo-Schroedinger}, while 
for the case of \(M=M_{p}\), this method is equivalent to a full configuration interaction approach.

To obtain the ground state within MCTHDB we rely on the so-called improved relaxation scheme.
This scheme can be summarized as follows: 
\begin{enumerate}
\item initialize the system with an ansatz set of single-particle functions \(\phi_i^{(0)}(r)\),
\item diagonalize the Hamiltonian within a basis spanned by the single-particle functions,
\item set the eigenvector with the lowest energy as the \(A^{(0)}\)-vector,
\item propagate the single-particle functions in imaginary time within a finite time interval \(d \tau\),
\item update the single-particle functions to \(\phi_k^{(1)} (r)\) and
\item repeat steps 2-5 until the state coefficients converge within the prescribed accuracy.
\end{enumerate}
For the diagonalization at step 2, the Lanczos approach is employed and for the propagation of \(\phi_k(r;\tau)\) at step 4, we employ the Dormand-Prince integrator.
For ensuring the consistency of the truncation with respect to \(M\), we have compared our \(M = 3\) and \(M = 4\) calculations verifying that the results presented in Fig. \ref{fig:comparisonMFMB} differ at most by \(0.1 \%\).
Our results shown in the main text correspond to \(M =4\).

\bibliographystyle{stylethesis}
\bibliography{bibliography_new}
\end{document}